\documentclass[letterpaper, twocolumn, journal]{IEEEtran}
\IEEEoverridecommandlockouts
\usepackage{cite}
\usepackage{amsmath, amssymb, amsfonts}
\usepackage{enumitem}
\usepackage{algorithmic}
\usepackage{graphicx}
\usepackage{subfigure}
\usepackage{textcomp}
\usepackage{xcolor}
\usepackage{amsthm}
\usepackage{multirow}
\usepackage{bm}
\usepackage{siunitx, nicematrix}
\usepackage{longtable, threeparttable, multirow}
\usepackage{makecell}

\usepackage{caption} 
\usepackage[font=small, labelfont=bf]{caption} 
\usepackage{hyperref}

\usepackage{soul}

\usepackage{booktabs}

\usepackage[]{hyperref} 
\hypersetup{
hidelinks,
colorlinks=false,
linkcolor=black,
citecolor=black,
}
\allowdisplaybreaks

\usepackage[ruled,vlined]{algorithm2e}
\SetKwInput{KwInput}{Input}  
\SetKwInput{KwOutput}{Output}

\newtheorem{lemma}{Lemma}
\newtheorem{theorem}{Theorem}

\newtheorem{corollary}{Corollary}

\newtheorem{example}{Example}

\newcommand{\Mod}[1]{\ (\mathrm{mod}\ #1)}
\newcommand\argmax{\mathop{\operatorname{arg\,max}}\limits}

\begin{document}

\title{Unified Block Signal Processing Framework for LPWANs: Sequence Index Modulation Spreading}



\author{Wenkun Wen, Tierui Min, Long Yuan, and Minghua Xia
	\thanks{The conference version of this paper was presented in the IEEE/CIC International Conference on Communications in China, Shanghai, China, Aug. 10-13, 2025 \cite{Wen-ICCC}.}
	\thanks{Wenkun Wen and Tierui Min are with the R\&D Department, Techphant Technologies Company Ltd., Guangzhou 510310, China (email: wenwenkun@techphant.net; mintierui@techphant.net)}
	\thanks{Long Yuan and Minghua Xia are with the School of Electronics and Information Technology, Sun Yat-sen University, Guangzhou 510006, China (email: yuanlong@mail2.sysu.edu.cn; xiamingh@mail.sysu.edu.cn)}
	}

\maketitle

\begin{abstract}
Low-power wide-area networks (LPWANs) demand high receiver sensitivity and efficient physical-layer signal processing. This paper introduces a unified framework for generalized block signal transmission in LPWANs, addressing the limitations of conventional symbol-by-symbol approaches. The framework comprises three key components: the signal block vector, the intra-block structure generator, and the signal basis matrix, and leverages quasi-orthogonal codewords formed through cyclically shifted spreading sequences. The resulting quasi-orthogonality enables reliable multi-user separation, particularly under asynchronous access. The framework establishes a conceptual foundation for block synchronization and provides a unified demodulation structure based on block correlation matching. It further supports flexible and systematic implementation, as demonstrated through applications to frequency-shift keying and chirp spread spectrum. This work advances scalable and efficient physical-layer design for next-generation LPWANs.
\end{abstract}


\section{Introduction}
As the Internet of Things (IoT) ecosystem continues to expand, low-power wide-area networks (LPWANs) have emerged as a key enabler for supporting massive deployments of energy-constrained devices. A central technical challenge in LPWAN design is achieving ultra-high receiver sensitivity while maintaining minimal power consumption and low hardware complexity. In practical systems, receiver sensitivity is often constrained by the performance of the synchronization module, which can become a critical bottleneck under harsh wireless propagation conditions.

Traditional wireless systems rely primarily on \textit{symbol-by-symbol} transmission models, which perform well in high signal-to-noise ratio (SNR) environments. Notable examples include Wi-Fi, 4G Long Term Evolution (LTE), and 5G New Radio (NR), which use quadrature amplitude modulation (QAM) in conjunction with precise synchronization techniques to achieve high-throughput communication. These systems depend on accurate recovery of the carrier frequency, phase, and symbol timing to prevent phase misalignment and inter-symbol interference. 

However, symbol-by-symbol approaches are often unsuitable for LPWAN applications, where devices operate under stringent energy constraints, low SNR conditions, and sporadic transmission patterns. Maintaining tight synchronization per symbol is challenging and can significantly degrade receiver sensitivity and reliability. This motivates the adoption of \textit{block-based} transmission frameworks that process signals in structured blocks, offering improved robustness to noise, interference, and timing offsets, while enabling low-complexity implementations suitable for energy-constrained IoT devices.

\subsection{Limitations of Symbol-by-Symbol Synchronization}
In conventional communication architectures, synchronization, detection, and decoding are performed on a per-symbol basis. These approaches are practical in moderately noisy channels with sufficient pilot symbols and signal power. Popular techniques include phase-locked loops and Costas loops for carrier recovery \cite{PT05}, as well as symbol timing recovery methods such as maximum-likelihood estimation, early-late gate tracking loops \cite{DCR98}, and classical algorithms like Gardner \cite{1096561} and Mueller-M\"{u}ller \cite{1093326}. While maximum-likelihood estimation is theoretically optimal, its performance deteriorates under practical non-idealities such as non-Gaussian noise, channel uncertainties, and model mismatches \cite{DC08}. Non-data-aided schemes, such as the Gardner algorithm, can perform adequately with oversampling but degrade significantly in the low-SNR regime or in time-varying channels, where noise and distortion impair timing recovery accuracy.

These challenges are amplified in LPWANs. Continuous synchronization loops are energy-intensive and unreliable in the ultra-low SNR regime (e.g., below $-20$~\si{dB}), where noise destabilizes the loop and causes frequent synchronization losses. Moreover, stringent power and complexity constraints render long pilot sequences or high-rate tracking impractical \cite{WenIoTJ25}. Symbol-by-symbol architectures also fail to accumulate energy across symbols coherently, a capability essential in low-SNR or fading channels \cite{Wen-ICCC}. These limitations have motivated growing interest in block-level signal processing as a remedy.

\subsection{Advantages of Block Transmission for Broadband Wireless Systems}
Block-based transmission models in broadband wireless communication systems were initially developed to mitigate multipath fading and inter-symbol interference when channel delay spread approaches the symbol duration. By grouping symbols into blocks and jointly processing them, systems can achieve more effective equalization, synchronization, and channel estimation.

Early work by Saltzberg \cite{Saltzberg1967} laid the foundation for multicarrier systems, later extended by Weinstein and Ebert \cite{Weinstein1971} using the Discrete Fourier Transform (DFT) to enable orthogonal frequency-division multiplexing (OFDM). The introduction of a cyclic prefix (CP) transformed linear convolution into circular one, simplifying frequency-domain equalization. Subsequent research \cite{Peled1980} enabled computationally efficient implementations, leading to widespread adoption in standards such as Digital Audio Broadcasting, Digital Video Broadcasting — Terrestrial, IEEE 802.11, 4G LTE, and 5G NR \cite{Bingham1990, 3gppLTE, 3gppNR}. Single-carrier block techniques, such as single-carrier frequency-domain equalization \cite{Sari1995, scfde, 5494776}, preserved the low peak-to-average power ratio of single-carrier systems while enabling frequency-domain equalization, and have been incorporated in uplink scenarios of 4G LTE and 5G NR \cite{lte_book, 5gnr}.

\subsection{Block Processing for LPWANs: A Missing Opportunity}
Despite employing long-duration waveforms, current LPWANs have not fully capitalized on the advantages of block-based processing. Technologies such as LoRa and Sigfox rely on long symbols or ultra-narrowband waveforms to achieve high sensitivity, yet their receivers still operate primarily through symbol-by-symbol demodulation supported by fragile synchronization loops. As a result, the underlying waveforms remain highly sensitive to phase noise, carrier frequency offsets, and time-varying channel conditions, especially in the low-SNR regimes typical of LPWAN deployments \cite{9000820, 9555814}.

Block-level processing offers several theoretical and practical advantages for LPWANs \cite{Wen-ICCC, YUAN25IoT}. Techniques like matched filtering, energy accumulation, and joint maximum-likelihood detection provide substantial \textit{processing gain}. For example, coherent accumulation over LoRa’s spreaded symbols can yield 20–30~\si{dB} SNR improvement relative to per-symbol detection. Doubling the spreading factor (SF) increases the gain by roughly 3~\si{dB}; for $\text{SF} = 7$, coherent despreading can deliver up to 21~\si{dB} improvement. Block detection schemes, including matched-filter and RAKE receivers, exploit total block energy, offering robustness against multipath and interference and outperforming conventional non-coherent schemes by 2.5–10~\si{dB} depending on the scenario \cite{Demeslay2022, Nguyen2021}. Nevertheless, adoption remains limited due to the absence of a unified design framework and concerns regarding implementation complexity.

\subsection{Contributions}  
While modern LPWANs leverage block transmission-like waveforms, such as LoRa's chirp spread spectrum and Sigfox's narrowband frames, their modulation and detection strategies are often ad hoc. This results in fragmented theoretical understanding and limited analysis of the performance-complexity tradeoff. This paper bridges these gaps by introducing a unified theoretical and architectural framework for block transmission in LPWAN physical layers. The main contributions include:

\begin{itemize}
	\item We introduce a generalized block transmission model for LPWANs, parameterized by a codebook space and an intra-block waveform generator, referred to as sequence index modulation spreading (SIMS). This model unifies various LPWAN modulations, including chirp and non-chirp schemes, under a common theoretical framework.
	\item We rigorously prove the quasi-orthogonality of distinct codewords within a SIMS codebook, as well as across multiple codebooks for different users, in a probabilistic sense.
	\item We propose a block-based multi-user transceiver architecture designed for synchronization, detection, and complexity management, optimized for energy- and bandwidth-constrained devices. This architecture provides a practical approach to implementing block-level processing on resource-limited hardware.
	\item We demonstrate how existing LPWAN schemes, such as LoRa and frequency-shift keying (FSK), can be interpreted and analyzed within the proposed framework. The effectiveness of this approach is validated through both theoretical analysis and simulation results, showing improvements in reliability, robustness, and multi-user scalability.
\end{itemize}

By moving beyond traditional per-symbol designs, this work introduces a new class of block-based LPWAN waveforms inherently robust to interference, noise, and synchronization errors. The framework provides a systematic foundation for designing flexible, resilient, and energy-efficient waveforms that support next-generation IoT ecosystems.

The remainder of this paper is organized as follows. Section~\ref{Section-Framework} develops a unified block signal processing framework for point-to-point transceivers. Section~\ref{Section-MultiUser} extends it to multi-user scenarios. Section~\ref{Section-Examples} illustrates the multi-user transceiver architecture with examples and comparisons to representative LPWAN technologies. Section~\ref{Section-Simulation} presents simulation results and discussion, followed by concluding remarks in Section~\ref{Section-Conclusion}.

{\it Notation:} Scalars, column vectors, and matrices are denoted by regular italic letters and lower- and upper-case letters in bold typeface, respectively. The superscripts $(\cdot)^T$, $(\cdot)^*$, and $(\cdot)^H$ denote the transpose, conjugate, and Hermitian transpose of a vector or matrix, respectively, and the subscript $\|\bm{x}\|_2$ denotes the $\ell_2$-norm of $\bm{x}$. The symbol $\bm{\mathsf{1}}$ indicates a column vector with all entries being unity, $\bm{I}$ represents an identity matrix of appropriate size, and $\mathrm{j} \triangleq \sqrt{-1}$ denotes the imaginary unit.  The abbreviation $\mathcal{CN} (\bm{0}, \, \varrho\bm{I})$ stands for a multi-variable complex Gaussian distribution with mean vector $\bm{0}$ and covariance matrix $\varrho \bm{I}$. The notation $\left| \mathcal{X} \right|$ denotes the cardinality of the set $\mathcal{X}$, and $\mathcal{Y} \setminus \mathcal{X}$ denotes the complement of set $\mathcal{Y}$ except for $\mathcal{X}$. The operator $\operatorname{diag}(\bm{x})$ denotes a diagonal matrix whose diagonal elements are given by the entries of $\bm{x}$ in the same order. The operator $\operatorname{vec}(\bm{X})$ indicates the vectorization of matrix $\bm{X}$ obtained by stacking its columns sequentially into a single column vector. The arithmetic operations $\bm{x} \succeq \bm{y}$, $\bm{x} \odot \bm{y}$, $\bm{x} \otimes \bm{y}$, $\bm{x} * \bm{y}$, and $\left\langle \bm{x}, \, \bm{y} \right\rangle$ denote that each element of $\bm{x}$ is greater than or equal to the counterpart of $\bm{y}$, the Hadamard product, the Kronecker product, the convolution, and the inner product of vectors $\bm{x}$ and $\bm{y}$, respectively. The Landau notation $\mathcal{O} ( \cdot )$ denotes the order of arithmetic operations, and the $Q$-function is defined as $Q(x) \triangleq \tfrac{1}{2\pi}\int_{x}^{\infty} {\exp\left(-{\mu^2}/{2}\right) \, {\rm d}\mu}$. Finally, for the reader’s convenience, the main symbols used throughout this paper are defined in Table~\ref{Tab_SymbolDefinition}.

   
    \begin{table}[t!]
    	\centering
    	\renewcommand\arraystretch{1.0}
    	\caption{Symobl Definition.}
    	\label{Tab_SymbolDefinition}
    	\begin{threeparttable}
    		\begin{tabular}{!{\vrule width0.12em} c | c !{\vrule width0.12em}}
    			\Xhline{0.12em} 
    			{\bf  Symbol} & \hspace{20pt}  {\bf Description} \hspace{20pt} \\
    			\Xhline{0.12em} 
			$\text{SF}$ & Spreading factor  \\
			\hline
			$E_s$ & Energy per symbol \\
			\hline
			$f_{0}$ & Subcarrier bandwidth \\
			\hline
			$K$ & Number of subcarries/Oversampling factor \\
			\hline
			$M$ & Number of users \\
			\hline
			$N_0$ & Thermal noise power spectral density \\
			\hline
			$N_b$ & Information bit block length \\
			\hline
			$N_\text{SF}$ & Spreaded block length \\
    			\Xhline{0.12em} 
    		\end{tabular}
	\vspace{-10pt}
    	\end{threeparttable}
    \end{table}

\section{Transceiver Architecture -- Unified Block Signal Processing and Waveform Design}
\label{Section-Framework}

\begin{figure}[!t]
    \centering
	\subfigure[Transmitter]{
		\includegraphics[width=0.375\textwidth]{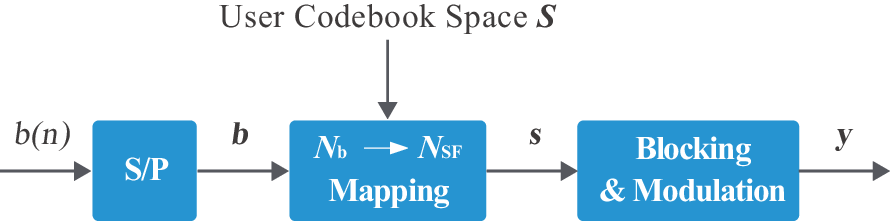} 
		\label{Fig_Tx}}	
	\vspace{-5pt}	
	\subfigure[Receiver]{
		\includegraphics[width=0.375\textwidth]{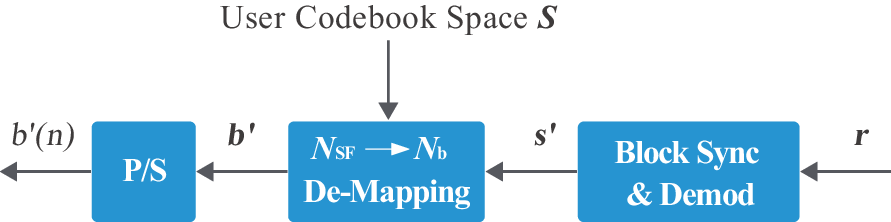}
		\label{Fig_Rx}}
	\vspace{-5pt}
	\caption{A general block signal processing transceiver for LPWANs.}
	\label{Fig_Transceiver}
	\vspace{-10pt}
\end{figure}

Fig.~\ref{Fig_Transceiver} presents the transceiver block diagram of a general block transmission system. At the transmitter, depicted in Fig.~\ref{Fig_Tx}, the input bitstream $b(n)$ is first converted from serial to parallel format, producing multiple information blocks, each of length $N_b$. Then, each block $\bm{b} \in \{0, 1\}^{N_b}$ of length $N_b$ is mapped to a codeword $\bm{s}$ of length $N_{\text{SF}}$ from a predefined spreading codebook, where $N_{\text{SF}} = 2^{\text{SF}}$ and $\text{SF}$ denotes the spreading factor. This process, generating a spreading code chip block $\bm{s}$, can be implemented via either a linear or a nonlinear mapping. It is important to note that typically $N_{\text{SF}} \gg N_b$; for instance, in LoRa's chirp spread spectrum, $N_{\text{SF}} = 2^{N_b}$. The resulting spreading code block $\bm{s}$ is then processed by a series of block operations, including modulation, to produce the transmit signal $\bm{y}$.

On the receiver side, as depicted in Fig.~\ref{Fig_Rx}, the received signal is first sampled to obtain the signal block $\bm{r}$. A synchronization module identifies the start of the relevant spreading block $\bm{s}^{\prime}$. The received block is demodulated and then despread through an inverse mapping operation, transforming the $N_{\text{SF}}$-length sequence back into an estimated information block $\bm{b}^{\prime} \in \{0, 1\}^{N_b}$. Finally, the recovered blocks are converted back to a serial bitstream $b^{\prime}(n)$ via a parallel-to-serial conversion. 

The block processing approach in Fig.~\ref{Fig_Transceiver} provides key advantages over symbol-by-symbol transmission, particularly in low-SNR and power-limited LPWANs:
\begin{itemize}
\item {\it Enhanced Synchronization and Demodulation}:
Long spreading codes applied across entire blocks improve correlation and energy per bit, enabling more accurate synchronization and robust demodulation under fading and interference. The adaptable spreading factor $\text{SF}$ allows flexible tradeoffs among rate, gain, and reliability.

\item {\it Parallel and Energy-Efficient Processing}:
Block structures support parallel fast Fourier transform (FFT), filtering, and matrix operations on CPUs, GPUs, and FPGAs, thereby improving throughput and latency while reducing the energy per operation, facilitating scalable, low-power implementations.

\item {\it Processing Gain and Noise Averaging}:
Spreading $N_b$ bits over $N_{\text{SF}}$ symbols yields a gain of $10\log_{10} N_{\text{SF}}$~\si{dB}, boosting noise and interference tolerance without bandwidth expansion. Block-level integration further enhances detection reliability in ultra-low-SNR regimes.
\end{itemize}

In summary, block processing improves synchronization, enables hardware parallelism, and enhances noise resilience, positioning it as a core technique for next-generation LPWANs under stringent energy and bandwidth constraints.

\subsection{Transmitter Block Signal Processing Framework}
This subsection introduces an inherent spreading processing framework at the transmitter, starting with the fundamental waveform.

\setul{}{1pt}

\subsubsection{\ul{Fundamental Modulation Waveform}}
For the general signal model of LPWANs, we define a complex-valued codebook space $\bm{S} \in \mathbb{C}^{K \times N_\text{SF}}$ as
\begin{align} \label{Eq-BW-Si}
	\bm{S} = [s(k, n)]_{K \times N_\text{SF}},
\end{align}
where $k = 0, \cdots, K-1$ and $n = 0, \cdots, N_\text{SF}-1$. Each entry $s(k, n)$ in the codebook is defined by
\begin{equation} \label{Eq-BW-Sin}
	s(k, n) = A(k) \cdot w(k, n),
\end{equation}
where $A(k) \in \mathbb{C}$ denotes a constant amplitude associated with each entry in the $k^\text{th}$ row of $\bm{S}$, and $w(k, n)$ is a fundamental modulation waveform specified by
\begin{equation} \label{Eq-w}
	w(k, n) = \exp\left( \mathrm{j} \frac{2 \pi}{K} f_0 \Lambda(k, n) \right), 
\end{equation}
with $f_0$ representing subcarrier bandwidth and $\Lambda(k, n)$ being a bivariate function of subcarrier index $k$ and time index $n$ that governs the intra-block waveform structure. The form of $\Lambda(k, n)$—linear, nonlinear, or piecewise—determines the internal characteristics of $\bm{S}$. For instance, a linear $\Lambda(k, n)$ corresponds to FSK or chirp spread spectrum (CSS) modulations, detailed in later sections. 


The fundamental waveform matrix $\bm{W} \in \mathbb{C}^{K \times N_\text{SF}}$ that spans the codebook space $\bm{S}$ is defined as
\begin{equation} \label{Eq-W}
	\bm{W} = \begin{bmatrix}
			w(0, 0) & \cdots & w(0, N_\text{SF}-1) \\
			\vdots & \ddots & \vdots \\
			w(K-1, 0) & \cdots & w(K-1, N_\text{SF}-1)
			\end{bmatrix}.
\end{equation}
Physically, each row of $\bm{W}$ corresponds to a frequency subcarrier or waveform component modulated over time via $\Lambda(k, n)$.

\subsubsection{\ul{Information Modulation Procudeure}}
In practical systems, each transmitted codeword 
\begin{equation}
	\bm{s}_k = \begin{bmatrix} s_k(0) & s_k(1) & \cdots & s_k(N_\text{SF}-1)\end{bmatrix}^H
\end{equation}
carries an information payload. The general modulation procedure is:

\begin{enumerate}[label={\it \alph*)}]
	\item {\it Codebook Construction}: Define the user-specific codebook space $\bm{S} = f(\bm{W}) = \begin{bmatrix}\bm{s}_0 & \bm{s}_1 & \cdots & \bm{s}_{K-1}\end{bmatrix}^H \in \mathbb{C}^{K \times N_\text{SF}}$, where $f(\cdot)$ denotes a scheme-dependent mapping applied to the fundamental waveform matrix $\bm{W}$.

	\item {\it Index Mapping}: Map each input bit block $\bm{b}_i = \begin{bmatrix}b_i (0) & b_i (1) & \cdots & b_i (N_b-1)\end{bmatrix}^T \in \{0, 1\}^{N_b}$ to an index $k = 0, \cdots, K-1$ via a deterministic mapping:
	\begin{align}
		k = {\rm Map}[\bm{b}_i],
	\end{align}
	where $b_i(n)$ is the $n^\text{th}$ bit of $\bm{b}_i$. For instance, one can take $k = \sum_{n=0}^{N_b - 1} b_i(n) \cdot 2^n$.

	\item {\it Codeword Selection}: Select $\bm{s}_k$ from $\bm{S}$ based on frequency index $k$, followed optionally by operations like CP-adding or zero-padding.

	\item {\it Waveform Modulation}: The final transmit waveform is modulated using FSK, CSS, or other compatible schemes with the structure of $\bm{W}$.
\end{enumerate}

This modular design enables flexible and robust block-level signal modulation for low-power wireless applications.

\subsection{Generalized Waveform Design: SIMS}
To generalize beyond linear $\Lambda(k, n)$ (as in FSK or CSS), we define
\begin{equation} \label{Eq-SIMS-Lambda}
	\Lambda(k, n) = \vartheta \bm{q}_l(n) n,
\end{equation}
where $\vartheta$ is a modulation index,\footnote{The modulation index $\vartheta$ is typically a fixed, scheme-dependent constant. For example, in continuous-phase frequency-shift keying (CPFSK), it is defined as $\vartheta = \Delta f_{\text{dev}} / B$, where $\Delta f_{\text{dev}}$ denotes the frequency deviation and $B$ the signal bandwidth. Within SIMS, however, $\vartheta$ takes on a more general role, serving as a scaling factor for arbitrary discrete spreading patterns.} and $k = \bm{q}_l(n) \triangleq \bm{q}_0[(n - l) \bmod N_\text{SF}]$ is the $n^\text{th}$ entry of the cyclically shifted root sequence $\bm{q}_0$, with shift $l = \text{Map}[\bm{b}_i]$. Substituting \eqref{Eq-SIMS-Lambda} into \eqref{Eq-BW-Sin}-\eqref{Eq-w} gives rise to a modulated spreading waveform:
\begin{equation} \label{Eq-BasicWaveform}
	s(k, n) = A(k) \cdot \exp\left( \mathrm{j} \frac{2 \pi}{K} f_0 \vartheta \bm{q}_l(n) n \right).
\end{equation}

The corresponding $l^\mathrm{th}$ codeword is defined as 
\begin{equation} \label{Eq-cl}
    \bm{c}_l = 
    \begin{bmatrix}
    \bm{g}_0^T & \bm{g}_1^T & \cdots & \bm{g}_{N_\text{SF}-1}^T
    \end{bmatrix}^T \in \mathbb{C}^{K N_\text{SF}},
\end{equation}
with 
\begin{equation} 
\label{Eq-gn}
    \bm{g}_n =
    \begin{bmatrix}
        \exp\left( \mathrm{j} \frac{2\pi}{K}  f_0 \vartheta \bm{q}_l(n) 0 \right) \\
        \exp\left( \mathrm{j} \frac{2\pi}{K}  f_0 \vartheta \bm{q}_l(n) 1 \right) \\
        \vdots \\
        \exp\left( \mathrm{j} \frac{2\pi}{K}  f_0 \vartheta \bm{q}_l(n) (K-1) \right)
    \end{bmatrix} \in \mathbb{C}^K,
    \end{equation}
	for all $n = 0, \cdots, N_\text{SF}-1$. Stacking the codewords $\bm{c}_l$, for all $l = 0, \cdots, N_\text{SF}-1$, yields the fundamental waveform matrix that spans the SIMS codebook space:
\begin{equation} 
	\bm{C} = f(\bm{W}) =
	\begin{bmatrix}
		\bm{c}_{0} & \bm{c}_{1} & \cdots & \bm{c}_{N_\text{SF}-1}
	\end{bmatrix}
	\in \mathbb{C}^{(K N_\text{SF}) \times N_\text{SF}}.
	\label{Eq-SIMS-W}
\end{equation}

We now demonstrate the Kronecker structure of SIMS-modulated codewords.

\begin{lemma}[Kronecker Structure of SIMS Codewords]
\label{Lemma-1}
Given a bit block $\bm{b}_i$ and its corresponding waveform samples $s(k, n)$ defined in \eqref{Eq-BasicWaveform}, the stacked codeword is obtained by column-wise vectorization as
\begin{equation}
    \bm{s}_l \triangleq \operatorname{vec}\!\left([\,s(k,n)\,]_{K \times N_{\mathrm{SF}}}\right) 
    \in \mathbb{C}^{K N_{\mathrm{SF}}},
\end{equation}
where $l=\mathrm{Map}[\bm{b}_i]$, $k=0,\ldots,K-1$, and $n = 0, \cdots,N_{\mathrm{SF}}-1$. 
This vector admits the Kronecker representation
\begin{equation} \label{eq:lemma1-main}
    \bm{s}_l
    = \left( \bm{I}_{N_{\mathrm{SF}}} \otimes \operatorname{diag}(\bm{d}_l) \right) \bm{c}_l,
\end{equation}
where $\bm{c}_l$ is defined in \eqref{Eq-cl}, and $\bm{d}_l$ denotes the vector of symbol amplitudes,
\begin{equation}
    \bm{d}_l
    = \begin{bmatrix}
        A(0) & A(1) & \cdots & A(K-1)
      \end{bmatrix}^{T}
      \in \mathbb{C}^{K}.
\end{equation}
\end{lemma}

\begin{proof}
	See the Appendix.
\end{proof}

The expression \eqref{eq:lemma1-main} decouples the \emph{symbol-dependent} vector $\bm{d}_l$ from the \emph{sequence-dependent} stack $\bm{c}_l$, a structural property that enables probabilistic cross-correlation bounds.

\begin{theorem}[Probabilistic Quasi-Orthogonality]
\label{Theorem-1}
Let the collections $\{\bm{s}_l\}_{l=0}^{N_\text{SF}-1}$ be SIMS codewords formed as in Lemma~\ref{Lemma-1}. For a fixed pair $0 \le i \neq j \le N_\text{SF}-1$, define the cross-correlation coefficient
\begin{equation}
	Z \triangleq \frac{1}{K N_{\rm SF}}\langle \bm{s}_i,\bm{s}_j\rangle.
\end{equation}
Assume that for each subcarrier index $k$, for all $k = 0, \cdots, K-1$, the per-subcarrier modulation symbols satisfy
\begin{equation}
	A_{\max} \triangleq \max_{k} \big|d^*_i[k] \, d_j[k]\big| < \infty,
\end{equation}
and that the per-subcarrier phase terms induced by spreading produce independent, zero-mean factors across subcarriers (standard under independently chosen spreading phases). Then for any $\epsilon>0$, it holds that
\begin{equation} \label{Eq:Theorem1-general}
	\Pr\big( |Z| \ge \epsilon \big)
		\le 2 \exp\!\left( -\frac{K N_{\rm SF}\epsilon^2}{2 A_{\max}^2 + \tfrac{2}{3} A_{\max}\epsilon} \right).
\end{equation}
\end{theorem}

\begin{proof}
Using the Kronecker representation of Lemma~\ref{Lemma-1}, the inner product admits the expansion
\begin{equation}
\langle \bm{s}_i, \bm{s}_j\rangle
	= \hspace{-0.5em} \sum_{n=0}^{N_{\rm SF}-1}\sum_{k=0}^{K-1}
		d^*_i[k] \, d_j[k] \exp\left(\mathrm{j}\frac{2\pi}{K} k f_0 \vartheta\big(c_i(n)-c_j(n)\big)\right).
\end{equation}
Define the elementary summands
\begin{equation}
	W_{k, n} \triangleq d^*_i[k] \, d_j[k] \exp\left(\mathrm{j}\frac{2\pi}{K} k f_0 \vartheta\big(c_i(n)-c_j(n)\big)\right).
\end{equation}
By hypothesis, the random phase factor has zero mean (for $k \neq 0$) and, together with independence across subcarriers, the collection \(\{W_{k,n}\}_{k,n}\) consists of independent, zero-mean complex random variables. Moreover, each summand is bounded in magnitude:
\begin{equation}
	|W_{k,n}| \le A_{\max}.
\end{equation}

Write the cross-correlation coefficient as
\begin{equation}
Z = \frac{1}{K N_{\rm SF}} \sum_{k, n} W_{k, n}.
\end{equation}
Equivalently consider the un-normalized sum \(S \triangleq \sum_{k, n} W_{k, n}\). Then \(Z = S/(K N_{\rm SF})\), and
\begin{equation}
	\Pr\big(|Z|\ge\epsilon\big) = \Pr\big(|S|\ge K N_{\rm SF}\epsilon\big).
\end{equation}

Let \(\sigma^2\) denote the total variance of \(S\):
\begin{equation} \label{Eq-Thm1-Proof-1}
	\sigma^2 = \sum_{k, n} \mathrm{Var}[W_{k, n}] \le \sum_{k, n} \mathbb{E}[|W_{k, n}|^2] \le K N_{\rm SF} A_{\max}^2.
\end{equation}
Applying Bernstein’s inequality \cite[Thm. 2.8.2]{Vershynin2025} separately to the real and imaginary parts of \(S\), with the uniform bound \(A_{\max}\) on the summands, for any \(t>0\), we have
\begin{equation}
\Pr\big(|S|\ge t\big)
\le 2 \exp\!\left( -\frac{t^2}{2\sigma^2 + \tfrac{2}{3} A_{\max} t} \right).
\end{equation}
Substituting \(t = K N_{\rm SF}\epsilon\) and the variance bound in \eqref{Eq-Thm1-Proof-1} yields
\begin{align}
\Pr\big(|Z|\ge\epsilon\big)
& \le 2 \exp\!\left( -\frac{K^2 N_{\rm SF}^2 \epsilon^2}
{2 K N_{\rm SF} A_{\max}^2 + \tfrac{2}{3} A_{\max} K N_{\rm SF} \epsilon} \right) \nonumber \\
&= 2 \exp\!\left( -\frac{K N_{\rm SF}\epsilon^2}{2 A_{\max}^2 + \tfrac{2}{3} A_{\max}\epsilon} \right),
\end{align}
which is the intended \eqref{Eq:Theorem1-general}.
\end{proof}

Theorem~\ref{Theorem-1} establishes a probabilistic bound on the cross-correlation between distinct SIMS codewords, showing that the likelihood of appreciable correlation decays exponentially with $K$ (the number of subcarriers) and $N_{\text{SF}}$ (proportional to the spreading factor). For sufficiently large $K$ and $N_{\text{SF}}$, the probability of harmful interference becomes negligible, ensuring robust quasi-orthogonality even under asynchronous access. These parameters introduce a natural performance--complexity tradeoff: increasing $K$ and $N_{\text{SF}}$ enhances multi-user separation and reliability, but also raises computational and bandwidth demands. This scaling behavior enables LPWAN systems to support large device populations while maintaining effective multi-user separation in noisy environments.

Regarding various constant-envelope modulation techniques widely used in LPWANs, we have the following corollary.

\begin{corollary}[Constant-Modulus Specialization]
\label{Corollary-Theorem1-const}
If the per-subcarrier modulation alphabet is constant-modulus so that \(|d_i[k]|=|d_j[k]|=1\) for all \(k = 0, \cdots, K-1\), then \(A_{\max}=1\) and \eqref{Eq:Theorem1-general} reduces to
\begin{equation}
	\Pr\big( |Z| \ge \epsilon \big)
	\le 2 \exp\!\left( - \frac{K N_{\rm SF} \epsilon^2}{2 + \tfrac{2}{3}\epsilon} \right).
\end{equation}
In particular, for moderate or large \(K N_{\rm SF}\), the tail probability decays exponentially fast in the product \(K N_{\rm SF}\).
\end{corollary}

The proposed SIMS framework offers several key improvements over conventional modulation schemes such as FSK and CSS, particularly in low-power and low-SNR LPWANs:
\begin{itemize}
	\item {\it Improved SNR Performance via Processing Gain:} SIMS inherently incorporates a spreading mechanism by associating each bit block with a cyclically shifted version of a pseudo-random root sequence. This design allows for constructive accumulation of signal energy across the block duration, resulting in a processing gain of approximately $10 \log_{10} N_\text{SF}$~\si{dB}. Compared to traditional FSK or CSS, where spreading is based on deterministic linear frequency variations, SIMS benefits from sequence diversity and enhanced correlation properties, yielding improved detection performance at low SNR.

	\item {\it Higher Spectral Efficiency Through Sequence Index Modulation:}  While classical FSK and CSS transmit one symbol per spreading block, SIMS uses {\it sequence index modulation} to encode multiple bits via the shift index $l = \text{Map}[\bm{b}_i]$. This enables the transmission of $\log_2 N_\text{SF}$ bits per spreading block, without increasing the bandwidth or sacrificing robustness \cite{YUAN25IoT}. Consequently, SIMS achieves higher spectral efficiency compared to baseline spreading schemes, making it suitable for bandwidth-constrained IoT applications.

	\item {\it Enhanced Resilience to Multipath and Interference:} The use of cyclically shifted pseudo-random sequences introduces favorable autocorrelation and cross-correlation properties, which improve resilience to time misalignment, multipath fading, and narrowband interference. This quasi-orthogonality (as formalized in Theorem~\ref{Theorem-1}) ensures that different codewords are distinguishable even under asynchronous reception, offering robustness that traditional linearly modulated CSS and FSK schemes lack in severely time-dispersive environments.

	\item {\it Increased Design Flexibility:} SIMS enables flexible waveform design by decoupling the spreading structure (via $\bm{c}_l$) from the modulation format (through $\vartheta$ and $\bm{d}_l$), as demonstrated in Lemma~\ref{Lemma-1}. This separation facilitates targeted optimization for application-specific requirements, such as ultra-low energy per bit, reduced peak-to-average power ratio, or simplified hardware implementation, making it particularly well-suited for LPWAN deployments.	
\end{itemize}

In summary, SIMS generalizes the traditional concept of spreading by introducing a structured index modulation layer within the codeword design. This enables significant improvements in both performance and resource efficiency, making SIMS a promising candidate for next-generation LPWAN physical-layer protocols.

\subsection{Receiver Block Processing Framework -- Joint Demodulation and Despreading}
Given the received signal 
\begin{equation}
	\bm{y} = \bm{s}_l + \bm{n},
\end{equation}
where $\bm{n}$ denotes a cyclically-symmetric additive Gaussian white noise (AWGN), it is necessary to find the codeword that best matches $\bm{y}$ from the SIMS codebook space $\bm{C}$ as the demodulated codeword, and then recover the information bits through the inverse mapping $N_\text{SF} \rightarrow N_{b}$. Thus, the typical coherent detection implies that
\begin{equation} \label{Eq-Detection-1} 
	\hat{l} = \argmax_{l = 0, 1, \cdots, N_{\text{SF}}-1} |\langle \bm{y}, \bm{c}_l \rangle|, \, \forall \bm{c}_l \in \bm{C}.
\end{equation}
Or more efficiently, \eqref{Eq-Detection-1} can be computed as
\begin{align}
	\bm{z} &= \bm{C}\bm{y}, \label{Eq-Detection-2a} \\
	\hat{l} &= \argmax_{l}|z_{l}|, \label{Eq-Detection-2b} 
\end{align}
where $z_{l}$ denotes the $l^{\text{th}}$ entry of $\bm{z}$.

In the context of SIMS, each codeword in the codebook $\bm{C}$ is constructed from a cyclically shifted version of a root sequence $\bm{q}_0$ of length $N_\text{SF}$. By organizing these shifted versions into a circulant matrix, denoted as $\bm{Q}(\bm{q}_0)$, the entire codebook space can be written compactly as
\begin{equation}
	\bm{C} = \bm{I}_K \otimes \bm{Q}(\bm{q}_0).
\end{equation}
This structure explicitly reveals that $\bm{C}$ consists of $K$ parallel circulant blocks, each corresponding to one phase state of the modulated waveform.

This inherent cyclicity enables a highly efficient receiver implementation. Specifically, the matched filtering step in \eqref{Eq-Detection-2a}, which requires a matrix-vector multiplication $\bm{C}\bm{y}$, can be replaced with FFT-based convolution, which diagonalizes circulant matrices in the frequency domain. As a result, the computational complexity is reduced from $\mathcal{O}\left(N_\text{SF}^2\right)$ to $\mathcal{O}\left(N_\text{SF} \log{N_\text{SF}}\right)$, meeting the stringent low-complexity and energy-efficiency requirements of LPWAN applications.

\section{Multi-User Scenarios}
\label{Section-MultiUser}
This section extends the point-to-point transceiver designed above to multi-user scenarios, including the codebook design and multi-user detection.

\subsection{Multi-User Codebook Spaces at The Transmitter}
In a multi-user SIMS system with $M$ users, each user is assigned a distinct spreading codebook denoted by $\bm{C}_{m}$, where $m = 0, 1, \cdots, M-1$. According to Lemma~\ref{Lemma-1}, these codebooks can be generated using different cyclic root sequences, such as $m$-sequences, Gold codes, or Zadoff-Chu sequences, to ensure quasi-orthogonality across users. The overall multi-user spreading codebook space is then formed by vertically concatenating the individual codebooks:
\begin{equation}
	\bm{C}_\text{MU} =
	\begin{bmatrix}
		\bm{C}_{0} \\
		\bm{C}_{1}\\
		\vdots \\
		\bm{C}_{M-1}
	\end{bmatrix}.
\end{equation}

Each user-specific codebook $\bm{C}_m$ is constructed by cyclically shifting a root sequence with good autocorrelation properties, resulting in (approximately) circulant structure. In practice, a single long pseudo-random sequence can be partitioned into distinct root segments to support scalable multi-user access with minimal degradation in cross-correlation performance.

The key enabler for low-interference multi-user communication lies in the quasi-orthogonality of codewords drawn from different users' codebooks. This ensures that inter-user interference remains bounded with high probability, even under asynchronous or overlapping transmissions.

\begin{theorem}[Probabilistic Quasi-Orthogonality Across Users]
\label{Theorem-2}
Let $\bm{c}_{i, \ell}$ and $\bm{c}_{j, m}$ denote the $i^{\mathrm{th}}$ and $j^{\mathrm{th}}$ SIMS codewords from users $\ell$ and $m$ with $\ell \neq m$. Suppose the spreading sums 
\begin{equation}
	D_l \triangleq \sum_{n=0}^{N_{\rm SF}-1} \exp\left(\mathrm{j}\frac{2\pi}{K} l f_0 \vartheta \left(c_{i, \ell}(n)-c_{j, m}(n)\right)\right)
\end{equation}
satisfy $|D_l| \le \mu N_{\rm SF}$, and the modulation products $d_{i, \ell}^*[n] d_{j, m}[n]$ are independent, zero-mean, and bounded by unit (constant-modulus alphabet). Then for any $\epsilon > 0$,
\begin{equation} \label{Eq-Thm2-prob}
	\Pr\!\left[\frac{1}{K N_{\rm SF}} \big|\langle \bm{s}_{i, \ell},\bm{s}_{j, m}\rangle\big| \ge \epsilon\right]
	\le 2 \exp\!\left(-\frac{K\epsilon^2}{2\mu^2 + \tfrac{2}{3}\mu\epsilon}\right).
\end{equation}
\end{theorem}

\begin{proof}
By Lemma~\ref{Lemma-1}, the inner product of $\bm{s}_{i, \ell}$ and $\bm{s}_{j, m}$ can be computed as
\begin{equation}
	\langle \bm{s}_{i, \ell},\bm{s}_{j, m} \rangle = \sum_{k=0}^{K-1} \big(d^*_{i, \ell}[k] \, d_{j, m}[k]\big) D_l.
\end{equation}
Using $|D_l|\le \mu N_{\rm SF}$, we obtain
\begin{equation} \label{Eq-Thm2-Proof-1}
	\frac{1}{K N_{\rm SF}}\big|\langle \bm{s}_{i, \ell},\bm{s}_{j, m} \rangle\big|
	\le \mu \left|\frac{1}{K}\sum_{k=0}^{K-1} d^*_{i, \ell}[k] \, d_{j, m}[k]\right|.
\end{equation}
The $K$ summands on the right-hand side of \eqref{Eq-Thm2-Proof-1} are independent, zero-mean, and bounded by unit in magnitude. Applying Bernstein's inequality \cite[Thm. 2.8.2]{Vershynin2025} separately to the real and imaginary parts yields
\begin{equation}
	\Pr\!\left[\left|\frac{1}{K}\sum_{k=0}^{K-1} d^*_{i, \ell}[k] \, d_{j, m}[k]\right| \ge \frac{\epsilon}{\mu}\right]
	\le 2 \exp\!\left(-\frac{K\epsilon^2}{2\mu^2 + \tfrac{2}{3}\mu\epsilon}\right),
\end{equation}
which implies the desired \eqref{Eq-Thm2-prob}.
\end{proof}

Leveraging the probabilistic quasi-orthogonality property of user codebooks, the residual multi-user interference is characterized in the following corollary.
\begin{corollary}[Residual Multi-User Interference]
\label{Corollary-2}
Let $\bm{s}_{i, \ell}$ be the desired codeword of user $\ell$ and suppose $M-1$ interferers transmit $\{\bm{s}_{j, m}\}_{\ell \neq m}$. If each interferer satisfies Theorem~\ref{Theorem-2} with violation probability:
\begin{equation} \label{Eq-Thm2-eta}
	\eta = 2 \exp\!\left(-\frac{K\epsilon^2}{2\mu^2 + \tfrac{2}{3}\mu\epsilon}\right),
\end{equation}
then with probability at least $1-(M-1)\eta$,
\begin{equation}\label{Eq:MUI-bound}
	\Big|\big\langle \bm{s}_{i, \ell}, \, \sum_{\ell \neq m}\bm{s}_{j, m} \big\rangle\Big|
	\le (M-1)\, K N_{\rm SF}\, \epsilon,
\end{equation}
and consequently, the power of residual multi-user interference is bounded by
\begin{equation}
	\Big|\big\langle \bm{s}_{i, \ell}, \, \sum_{\ell \neq m}\bm{s}_{j, m} \big\rangle\Big|^2
	\le (M-1)^2 (K N_{\rm SF})^2 \epsilon^2.
\end{equation}
\end{corollary}

In Theorem~\ref{Theorem-2}, it is assumed that $|D_l|\le \mu N_{\rm SF}$, where $\mu$ represents the normalized cross-correlation of user spreading sequences. By the Welch bound \cite{6151774}, the minimum achievable value of $\mu$ for $M$ unit-norm vectors $\{\bm{v}_i\}_{i=1}^{M}$ in $\mathbb{C}^{N_{\rm SF}}$, with $M \gg N_{\rm SF}$, is lower bounded as
\begin{equation} \label{Eq-WelchBound}
	\max_{i \ne j} |\langle \bm{v}_i, \bm{v}_j \rangle| \ge \sqrt{\frac{M-N_{\rm SF}}{N_{\rm SF}(M-1)}},
\end{equation}
which is approximately attained by well-designed spreading families such as $m$-sequences, Gold codes, or Zadoff-Chu sequences. Hence, in practice, $\mu \approx 1/\sqrt{N_{\rm SF}}$, and increasing $N_{\rm SF}$ tightens the bound in Theorem~\ref{Theorem-2}. For asynchronous multi-user operation, it is therefore recommended to select a sequence family with coherence parameter $\mu \approx 1/\sqrt{N_{\text{SF}}}$, choose $N_{\rm SF}$ large enough to achieve the desired confidence parameter $\eta$ for a given tolerance $\epsilon$, and then apply~\eqref{Eq-Thm2-eta} to trade off $K$, $N_{\rm SF}$, and $\epsilon$ during system design.

In practice, the quasi-orthogonality guarantees in Theorems~\ref{Theorem-1} and \ref{Theorem-2} depend on the joint selection of spreading sequences and modulation schemes:
\begin{itemize}
    \item {\it Spreading Sequences}: Classical families such as $m$-sequences and Gold codes are commonly used due to their favorable autocorrelation and low cross-correlation properties. Gold codes, in particular, offer greater flexibility in codebook size while maintaining near-Welch-bound coherence, which is particularly advantageous for multi-user deployments.
    \item {\it Modulation Schemes}: Phase-shift keying (PSK) ensures constant-modulus symbols, resulting in $A_{\max}=1$ in Theorem~\ref{Theorem-1}, thus maximizing the strength of the concentration bound. Continuous phase modulation (CPM), although more complex to implement, provides excellent spectral compactness and robustness to nonlinearities while also satisfying the boundedness and zero-mean requirements of our analysis.
\end{itemize}

This hierarchical design leverages the flexibility of the SIMS framework and the efficiency of block processing, offering a scalable, robust, and low-power solution for multi-user access in LPWAN environments.

\subsection{Multi-User Detection at The Receiver}
At the receiver, joint detection of multiple users can be efficiently performed by leveraging the quasi-orthogonality between user-specific spreading codebooks. Let $\bm{y}$ denote the received signal comprising the superposition of all $M$ users:
\begin{equation}
	\bm{y} = \sum_{m=0}^{M-1} \bm{s}_{j_m, m} + \bm{n},
\end{equation}
where $\bm{s}_{j_m, m}$ is the transmitted codeword of the $m^\text{th}$ user, and $\bm{n}$ is an AWGN. Note that the subscript $j_m$ of $\bm{s}_{j_m, m}$ indicates that the codeword index $j_m$ varies with the user index $m$.

To jointly detect the transmitted codewords, we compute the matched filtering outputs over the entire multi-user codebook space $\bm{C}_{\text{MU}}$:
\begin{align}
	\bm{z} &= \bm{C}_\text{MU} \bm{y}, \label{Eq-MU-Detect-a} \\
	\hat{\jmath}_m &= \argmax_{k} \left|z_{j}^{(m)}\right|, \, m = 0, \cdots, M-1, \label{Eq-MU-Detect-b}
\end{align}
where $z_j^{(m)}$ denotes the correlation output between the received signal and the $j^\text{th}$ codeword in the $m^\text{th}$ codespace $\bm{C}_{m}$.

Thanks to the quasi-orthogonality between users’ codebooks (as shown in Theorem~\ref{Theorem-2}), inter-user interference remains minimal even under simultaneous transmissions. This allows near-optimal detection by evaluating inner products in parallel for each user's codebook independently, without requiring successive interference cancellation or complex joint decoding.

Moreover, if the codebooks $\bm{C}_{m}$ are designed as circulant matrices, the detection process \eqref{Eq-MU-Detect-a}--\eqref{Eq-MU-Detect-b} can be accelerated via FFT and IFFT operations:
\begin{equation}
	\bm{z}^{(m)} = \text{IFFT}\left\{ (\text{FFT}(\bm{q}_m))^* \odot \text{FFT}(\bm{y}) \right\}, \, \forall m,
\end{equation}
where $\bm{q}_m$ is the root sequence used to generate $\bm{C}_m$.

The computational complexity of multi-user detection scales linearly with the number of users $M$, and as $\mathcal{O}\left(N_\text{SF} \log{N_\text{SF}}\right)$ per user due to FFT-based matched filtering. Therefore, the total complexity remains manageable for a large number of users and is well-suited for implementation on parallel hardware (e.g., GPUs or FPGAs). This scalability advantage makes SIMS a promising candidate for dense IoT environments with massive device connectivity requirements.

\section{Multi-User Transceiver Architecture, Illustrating Examples, and Comparison}
\label{Section-Examples}
In this section, we present the multi-user transceiver architecture of the SIMS, followed by two examples using the generalized block signal model described above, which include the typical FSK and CSS modulations for illustrative purposes. Finally, we make a comprehensive comparison with representative LPWAN technologies.

\subsection{Multi-User Transceiver Architecture}
\begin{figure*}[!t]
        	\centering
        	\includegraphics[width=0.7\textwidth]{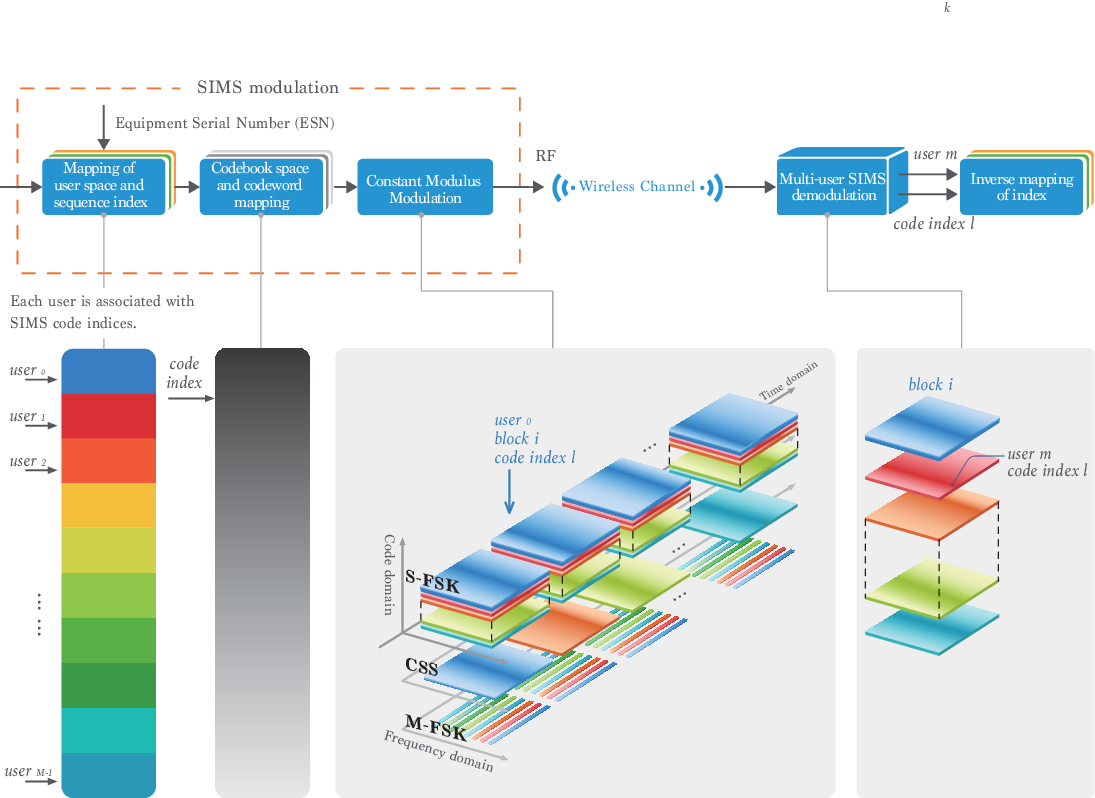} 
	\caption{Multi-user transceiver architecture of the SIMS system. The transmitter selects a user-specific spreading codeword from a structured codebook and applies constant-modulus modulation. The receiver uses FFT-based block processing for synchronization and matched filtering to recover user data.}
        	\label{Fig_TxRx}
	\vspace{-10pt}
\end{figure*}

Fig.~\ref{Fig_TxRx} presents the multi-user transceiver architecture. On the transmitter side, the system operates in three stages: codebook space assignment, spreading codeword selection, and constant-modulus modulation. Each user $m = 0, 1, \cdots, M-1$ is allocated a unique subspace $\bm{C}_m$ within the global multi-user codebook $\bm{C}_\text{MU}$, determined by its equipment serial number (ESN). The information bit block of length $N_b$ is mapped to a codeword index $l = \text{Map}[\bm{b}_i]$ inside the assigned subspace. The selected spreading codeword is then modulated using a constant-envelope waveform and transmitted over the channel.

At the receiver, the ESN is identified during synchronization, enabling the system to retrieve the correct codebook $\bm{C}_m$ for each user. Block detection then follows \eqref{Eq-MU-Detect-a} and \eqref{Eq-MU-Detect-b}, allowing for reliable data recovery via FFT-based matched filtering.

\subsubsection{\ul{Synchronization}}

The SIMS synchronization scheme offers notable advantages over synchronization mechanisms in mainstream LPWAN standards, such as LoRa, Sigfox, and NB-IoT, providing superior timing accuracy, low-SNR robustness, and hardware efficiency. A concise comparison is given below:

\begin{itemize}
\item {\it LoRa}: Synchronization is achieved via time-domain correlation between received chirps and reference chirps \cite{9000820}. Despite strong noise immunity \cite{WenIoTJ25}, this serial implementation introduces latency and is tied to chirp waveforms. In contrast, SIMS performs block-level synchronization using cyclic pilots and FFT-based parallel correlation, yielding faster processing and flexible spreading sequences (e.g., PN or Zadoff–Chu).

\item {\it Sigfox}: Using ultra-narrowband signals, Sigfox relies on coarse time-of-arrival detection with centralized processing and long detection windows. This reduces device complexity but increases latency and degrades performance in the presence of mobility or fading. SIMS instead supports fully decentralized synchronization with lightweight local processing and improved tolerance to multipath and frequency offsets.

\item {\it NB-IoT}: Based on OFDM synchronization with the primary synchronization signal and the secondary synchronization signal \cite{11017626}, NB-IoT achieves high robustness but at significant power and complexity costs, unsuitable for ultra-low-power nodes. SIMS offers comparable resilience with lower complexity through FFT-based $\mathcal{O}\left(N_\text{SF}\log\left(N_\text{SF}\right)\right)$ processing, thereby avoiding the tight timing alignment and coordination overhead.
\end{itemize}



%
\subsubsection{\ul{Multi-User Access Strategies}} 

Table~\ref{tab-mu-comparison} summarizes representative multi-user access strategies in LPWANs. Unlike LoRa, which separates users through time–frequency allocation via spreading factors, SIMS achieves user separation entirely in the code domain using quasi-orthogonal cyclic codebooks. This enables massive user scalability without centralized scheduling, while maintaining low receiver complexity through FFT-based correlation. Sigfox employs ultra-narrowband random ALOHA transmissions, which offer minimal complexity but limited scalability and poor interference tolerance. NB-IoT relies on tightly scheduled time–frequency resources and incurs substantial signaling overhead, whereas SIMS offers a lightweight, distributed alternative. Overall, SIMS combines high concurrency, robustness, and computational efficiency, positioning it as a strong candidate for future massive-access LPWANs.

\begin{table*}[!t]
	\centering
	\caption{Comparison of Multi-User Access Strategies in LPWANs}
	\label{tab-mu-comparison}
	\begin{tabular}{lcccc}
		\toprule
		\textbf{Technology} & \textbf{Access Scheme} & \textbf{User Separation} & \textbf{Scalability} & \textbf{Receiver Complexity} \\
		\midrule
		\textbf{SIMS (proposed)} & Code Division & Quasi-orthogonal cyclic codebooks & High & Low ($\mathcal{O}(N_\text{SF} \log N_\text{SF})$) \\
		\textbf{LoRa} & TDMA/FDMA & SF and channel & Moderate & Low (chirp correlation) \\
		\textbf{Sigfox} & ALOHA & Frequency/time diversity (randomized)  & Low--Moderate & Very Low (energy detection) \\
		\textbf{NB-IoT} & Scheduled TDMA/FDMA & Centralized allocation & Limited & Medium--High \\
		\bottomrule
	\end{tabular}
\end{table*}

\setcounter{equation}{50}
\begin{figure*}[!t]
	\normalsize
	\vspace*{4pt}
	\begin{align}
		\bm{C}_\text{CSS} = f(\bm{W}) = \begin{bmatrix}
			w(0, 0) & w(1, 1) & \cdots & w(N_\text{SF}-1, N_\text{SF}-1) \\
			w(1, 1) & w(2, 2) & \cdots & w(0, 0) \\
			 \vdots & \vdots & \ddots & \vdots \\
			w(N_\text{SF}-1, N_\text{SF}-1) & w(0, 0) & \cdots & w(N_\text{SF}-2, N_\text{SF}-2)
		\end{bmatrix}
		\label{Eq-CSS-S}
		\end{align}
	\hrulefill
\end{figure*}
\setcounter{equation}{45}

\subsection{Two Illustrating Examples}
\begin{example}[FSK Modulation] 
\label{Ex-MFSK}
By using the generalized block signal model above, the conventional FSK modulation can be obtained as follows.
\begin{itemize}
	\item Establish codebook space: The spreading codebook space of FSK is essentially an IDFT matrix of size $K \times N_\text{SF}$, as shown in \eqref{Eq-W}, and each column represents a $K$-ary FSK signal waveform, such that we have
\begin{equation} \label{Eq-MFSK-S}
	\bm{C}_\text{\rm FSK} = f(\bm{W}) = \bm{W}.
\end{equation}
In this case, the parameter $K$ denotes the number of frequency subcarriers employed in the FSK modulation scheme.
	\item Map codeword index: Map $N_b = \log_2{K}$ bits of information to a decimal index $k$.
	\item Select the codeword: Select the $k^{\rm th}$ codeword from the codebook space $\bm{C}_\text{\rm FSK}$ by index $k$, and the Tx codeword is constructed by
\begin{align}
	\bm{s}_k 
		&= A_k \cdot \left(\bm{W}[k, :]\right)^H  \nonumber \\
		&= A_k \cdot \begin{bmatrix} w(k, 0) & w(k, 1) & \cdots & w(k, N_\text{SF}-1)\end{bmatrix}^H,
\end{align}
where $\bm{W}[k, :]$ represents the $k^{\rm th}$ row of the basis matrix $\bm{W}$, whose $(k, n)^{\rm th}$ entry is explicitly given by 
\begin{align} \label{Eq-MFSK-w}
	w(k, n) = \exp\left( \mathrm{j} \frac{2\pi}{N_\text{SF}} f_0 \vartheta k n \right),
\end{align}
which, compared to \eqref{Eq-w}, implies the phase $\Lambda(k, n)$ is linearly increases with $k$, i.e., 
\begin{equation} \label{Eq-MFSK-Lambda}
	\Lambda(k, n) = \vartheta k n.
\end{equation}
\end{itemize}
\end{example}

\begin{example}[CSS Modulation]
\label{Ex-CSS}
This example illustrates how to generate CSS signals using the generalized block signal model above.
\begin{itemize}
	\item Map codeword index: Map $N_b = \log_2{N_\text{SF}} = \text{SF}$ bits of information to a decimal index $m$. Note that the values of $\text{SF}$ is limited to the set $\{7, 8, \cdots, 12\}$ in LoRa modulation but has no limit in our framework.
	\item Establish codebook space: The entry $c(k, n)$ of the CSS codeword $\bm{c}_k$ is taken from the diagonal entries determined by $n$ and $m+n \Mod{N_\text{SF}}$ in $\bm{W}$  \cite{8723130}: 
\setcounter{equation}{49}
\begin{equation} \label{Eq-50}
	c(k, n) = A(k) \cdot w\left(m+n \Mod{N_\text{SF}}, n\right).
\end{equation}
\setcounter{equation}{51}
Clearly, the codeword $\bm{c}_k$ is a column vector composed of the diagonal entries of $\bm{W}$, including all cyclic diagonals parallel to the main diagonal. As a result, the CSS basis matrix $\bm{C}_{\mathrm{CSS}}$ takes the form shown in \eqref{Eq-CSS-S} at the top of the page, where each column is obtained by a cyclic shift of its predecessor. In comparison with \eqref{Eq-w}, this structure indicates that the frequency index $k$ increases linearly with the time index $n$, leading to
\begin{equation} \label{Eq-CSS-Lambda}
    \Lambda(k,n) = \vartheta n^{2}.
\end{equation}

	\item Select the codeword: select the $k^\text{\rm th}$ codeword (which is already modulated by recalling \eqref{Eq-50}) from the codebook space $\bm{C}_\text{\rm CSS}$ by the index $k$.
\end{itemize}
\end{example}

When comparing the fundamental waveform matrix $\bm{C}$ shown in \eqref{Eq-SIMS-W} with $\bm{C}_\text{FSK}$ as defined in \eqref{Eq-MFSK-S} for FSK modulation, or $\bm{C}_\text{CSS}$ as shown in \eqref{Eq-CSS-S} for CSS modulation, it can be observed that those for both FSK and CSS are fixed. In contrast, the fundamental waveform matrix $\bm{C}$ for SIMS can be flexibly defined by selecting a different root random sequence $\bm{q}_{0}$. This flexibility enables the creation of various SIMS waveform matrices, resulting in significant benefits for applications with a large number of users. 

\subsection{Comparison with Representative LPWAN Technologies}
As shown in Table~\ref{Tab-Codebook-Extended}, SIMS generalizes the traditional block-based modulations by allowing dynamic construction of the codebook space through varying root sequences. While FSK and CSS use fixed waveforms derived from the IDFT matrix, SIMS supports a diverse, configurable set of quasi-orthogonal waveforms. This enhances flexibility, spectral efficiency, and scalability in multi-user LPWANs.

\begin{table*}[!t]
	\centering
	\caption{Comparison of FSK, CSS, and SIMS Modulation Schemes}
	\label{Tab-Codebook-Extended}
	\begin{tabular}{lccc}
		\toprule
		\textbf{Feature} & \textbf{M-FSK} & \textbf{CSS} & \textbf{SIMS} \\
		\midrule
		Codebook structure & Rows of IDFT matrix & Cyclic diagonals of IDFT & Root-sequence-based construction \\
		Codebook flexibility & Fixed & Semi-flexible & Highly flexible \\
		Codebook size & $K$ & $N_\text{SF}$ & Configurable \\
		Root sequence required & No & No & Yes \\
		Orthogonality level & Full (ideal) & Partial (cyclic) & Near-orthogonal (randomized) \\
		Spreading gain & Low (typically 1) & High ($N_\text{SF}$) & Configurable via $K, N_\text{SF}$ \\
		SNR efficiency & Moderate & Good & Excellent (via spreading and coding) \\
		Resilience to interference & Low & Medium (cyclic shift diversity) & High (sequence diversity + block processing) \\
		Implementation complexity & Low & Moderate & Moderate to High \\
		Multi-user scalability & Poor & Limited & Strong (codebook division) \\
		Spectral efficiency & Low & Medium & Medium to High \\
		Synchronization overhead & Minimal & Preamble-based & Integrated block-wise correlation \\
		Standard usage & Sigfox & LoRa & New (SIMS) \\
		\bottomrule
	\end{tabular}
\end{table*}

To highlight the unique capabilities of the proposed SIMS framework, we compare its key features with several widely adopted LPWAN technologies, including LoRa, Sigfox, and NB-IoT. Table~\ref{Tab-PhysicalLayer} summarizes the differences at the physical layer, including modulation formats, bandwidth usage, data rates, and synchronization strategies. SIMS offers a flexible tradeoff between spectral efficiency and robustness by leveraging block-based CPM and blind synchronization. In contrast, traditional schemes, such as LoRa and Sigfox, rely on fixed waveform structures and explicit preambles.

\begin{table*}[!t]
	\centering
	\caption{Physical Layer Comparison of LPWAN Schemes}
	\label{Tab-PhysicalLayer}
	\begin{tabular}{lcccc}
	\toprule
	\textbf{Feature} & \textbf{LoRa} & \textbf{Sigfox} & \textbf{NB-IoT} & \textbf{SIMS} \\
	\midrule
	Modulation & CSS & BPSK/DBPSK & QPSK (OFDM/SC-FDMA) & CPM-based block modulation \\
	Bandwidth & 125–500 kHz & 100 Hz & 180 kHz & 10–500 kHz (flexible) \\
	Data Rate & 0.3–50 kbps & 100 bps & Up to 250 kbps & Flexible (codebook-defined) \\
	Spectral Efficiency & Low & Very low & High & Medium to high \\
	Synchronization & Preamble & Pattern detect & Frame-based & Code-aligned (blind, robust) \\
	\bottomrule
	\end{tabular}
\end{table*}

\section{Simulation Results and Discussions}
\label{Section-Simulation}

In this section, we evaluate the performance of the proposed SIMS framework and compare it with representative LPWAN modulation schemes, including M-ary FSK (M-FSK) and CSS, under various channel and system conditions. All simulations are conducted using baseband discrete-time models, and the results are averaged over $2 \times 10^6$ Monte Carlo trials.

\subsection{Single-User Scenarios}
To verify the theoretical processing gain achieved by block-based spreading, we simulate and compare the bit-error-rate (BER) performance of M-FSK and CSS over AWGN, Rayleigh, and frequency-selective fading channels. All schemes are normalized to have equal energy per bit. 

\subsubsection{\ul{AWGN Channels}}
The BER of noncoherent detection of orthogonal signals over AWGN channels is given by \cite[Eq.~(4.5-44)]{DC08}:
\begin{equation} \label{Eq-BER-AWGN-accurate}
	P_b = \frac{2^{\text{SF}-1}}{2^{\text{SF}}-1} \sum_{k=1}^{2^{\text{SF}}-1} \frac{(-1)^{k+1}}{k+1} \binom{2^{\text{SF}}-1}{k} e^{-\frac{k}{k+1} \Gamma},
\end{equation}
where $\Gamma=E_s/N_0$.

Although \eqref{Eq-BER-AWGN-accurate} is exact and analytical, for large modulation orders (equivalently, large $\text{SF}$), the binomial terms may cause numerical instability in MATLAB implementations \cite{BER-LoRa}. To ensure stability, the following closed-form approximation can be used \cite[Eq.~(23)]{BER-LoRa}:
\begin{equation} \label{Eq-BER-AWGNApprox}
	P_b \approx 0.5 \times Q\left(\sqrt{\Gamma \times 2^{\text{SF}+1}} - \sqrt{1.386 \times \text{SF} + 1.154}\right).
\end{equation}

\begin{figure}[!t]
	\centering
	\includegraphics [width=0.35\textwidth]{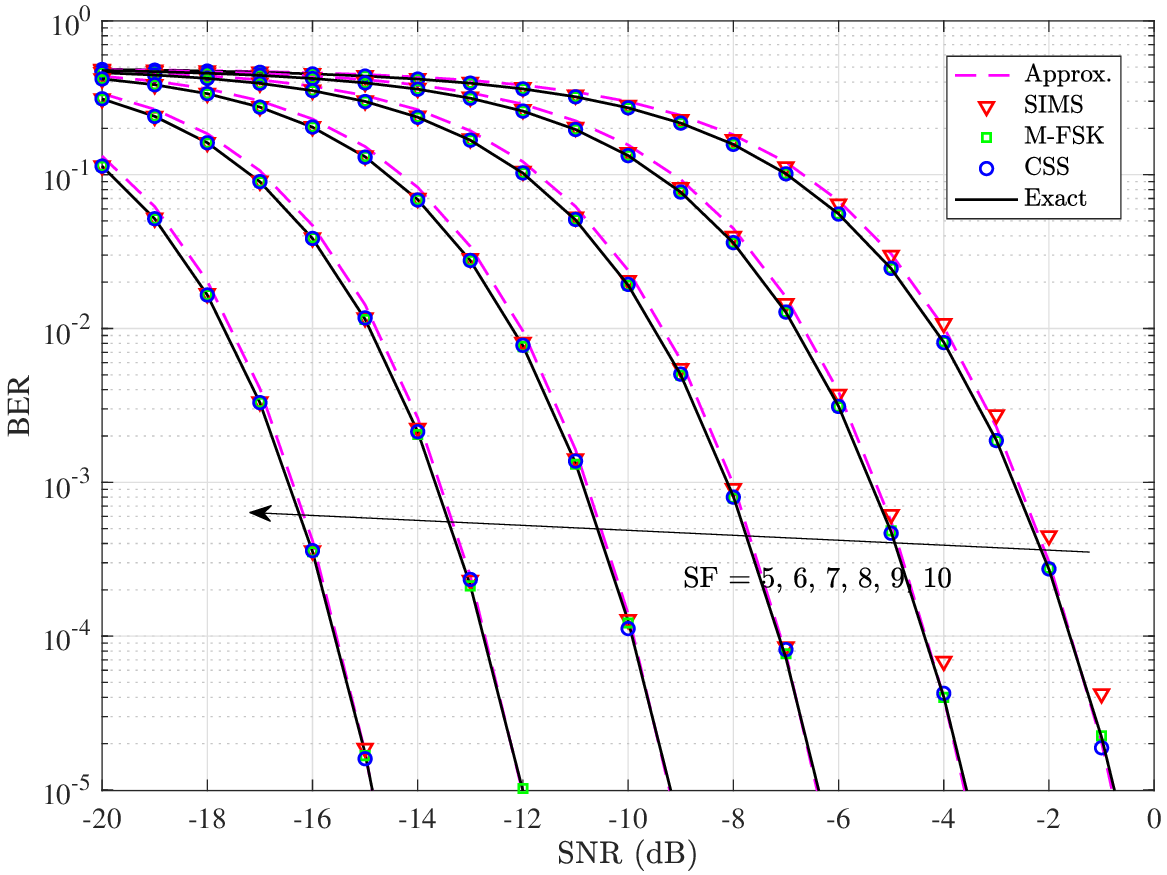}
	\caption{BER performance comparison of SIMS, CSS, and M-FSK schemes over AWGN channels for different spreading factors.}
	\label{Fig_BER_AWGN}
	\vspace{-10pt}
\end{figure}

Fig.~\ref{Fig_BER_AWGN} shows the BER comparison of SIMS, CSS, and M-FSK in AWGN channels for different SF values. The exact and approximate BERs are computed by \eqref{Eq-BER-AWGN-accurate} and \eqref{Eq-BER-AWGNApprox}, respectively. As expected, the approximate BERs deviate slightly from the exact values, while the simulated BERs closely match the same results. For small SF values (e.g., $\text{SF}=5, 6$), the SIMS scheme exhibits slightly higher BERs than CSS and M-FSK due to its use of quasi-orthogonal sequences. However, as SF increases, the BER performance of SIMS converges to that of CSS and M-FSK, consistent with Theorem~\ref{Theorem-1}. Moreover, each increment in SF yields roughly a $3$ \si{dB} gain, corresponding to the expected spreading gain from doubling the sequence length.

\subsubsection{\ul{Rayleigh Fading Channels}}
The BER of noncoherent detection of orthogonal signals over Rayleigh fading channels is given by \cite[Eq.~(13.4-50)]{DC08}:
\begin{equation} \label{Eq-BER-Rayleigh-accurate}
	P_b = \frac{2^{\text{SF}-1}}{2^{\text{SF}}-1} \sum_{k=1}^{2^{\text{SF}}-1} \frac{(-1)^{k+1}\binom{2^{\text{SF}}-1}{k}}{1+k+k \overline{\Gamma}}, 
\end{equation}
where $\overline{\Gamma} = \mathbb{E}(|h|^2) E_s / N_0$ and $h \sim \mathcal{CN}(0,1)$.

Similar to the approxiamtion of \eqref{Eq-BER-AWGNApprox} to \eqref{Eq-BER-AWGN-accurate}, for large $\text{SF}$ a numerically stable approximation to \eqref{Eq-BER-Rayleigh-accurate} can be applied \cite[Eq.~(33)]{BER-LoRa}:
\begin{align} \label{Eq-BER-Rayleigh-Approx}
     	P_b & \approx 0.5 \times \left[Q\left(-\sqrt{2H_{2^{\text{SF}}-1}}\right) - \sqrt{\frac{\Gamma_\text{eff}}{\Gamma_\text{eff}+1}} \exp\left(-\frac{H_{2^{\text{SF}}-1}}{\Gamma_\text{eff}+1}\right) \right.  \nonumber \\
	& \quad \left. \times Q\left(\sqrt{\frac{\Gamma_\text{eff}+1}{\Gamma_\text{eff}}} \left(-\sqrt{2H_{2^{\text{SF}}-1}}+\frac{\sqrt{2H_{2^{\text{SF}}-1}}}{\Gamma_\text{eff}+1}\right)\right)\right], 
\end{align}
where $H_n \triangleq \sum_{k=1}^n {1/k}$ denotes the $n^\text{th}$ harmonic number, and $\Gamma_\text{eff} \triangleq 2^{\text{SF}} E_s/N_0$. For large $n$, $H_n \approx \ln(n) + \tfrac{1}{2n} + 0.57722$, with $0.57722$ being the Euler–Mascheroni constant.

\begin{figure}[!t]
	\centering
	\includegraphics [width=0.35\textwidth]{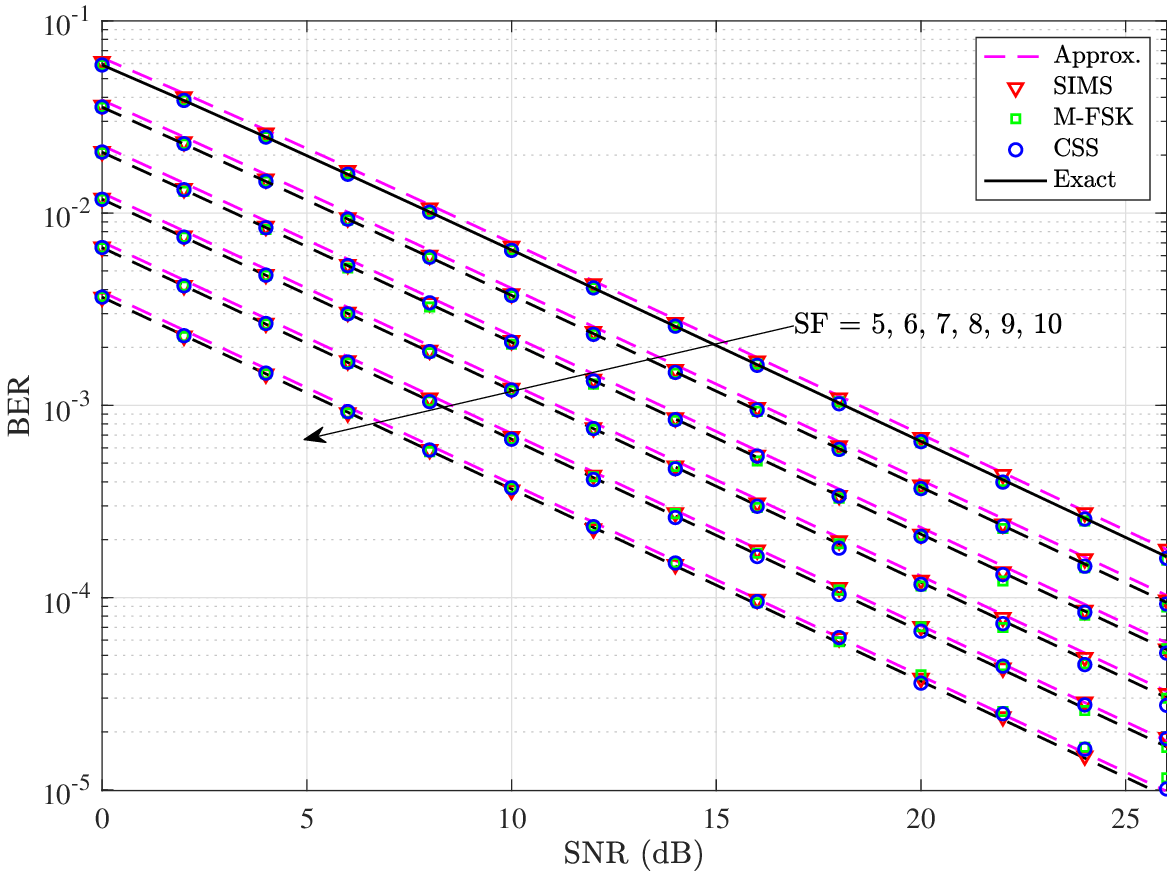}
	\caption{BER performance comparison of SIMS, CSS, and M-FSK schemes over Rayleigh channels for different spreading factors.}
	\label{Fig_BER_Rayleigh}
	\vspace{-10pt}
\end{figure}

Fig.~\ref{Fig_BER_Rayleigh} presents the BER results for Rayleigh fading channels, showing trends similar to those observed in Fig.~\ref{Fig_BER_AWGN} for AWGN channels. Both figures confirm the feasibility of the SIMS scheme and validate the accuracy of the BER analyses.

\subsubsection{\ul{Frequency-Selective Fading Channels}}
To evaluate performance under multipath conditions, we consider a frequency-selective two-tap channel with impulse response:
\begin{equation} \label{Eq-FSCM}
    h(n) = \sqrt{1-\rho} \, \delta(nT) + \sqrt{\rho} \, \delta(nT-T),
\end{equation}
where $T$ is the sampling duration. This model has been widely adopted in the literature (e.g., \cite{FSCM-LoRa, DM-TDM-CSS}). In the simulations, the power-splitting factor is set to $\rho = 0.2$ in \eqref{Eq-FSCM}. 

\begin{figure}[!t]
	\centering
	\includegraphics[width=0.35\textwidth]{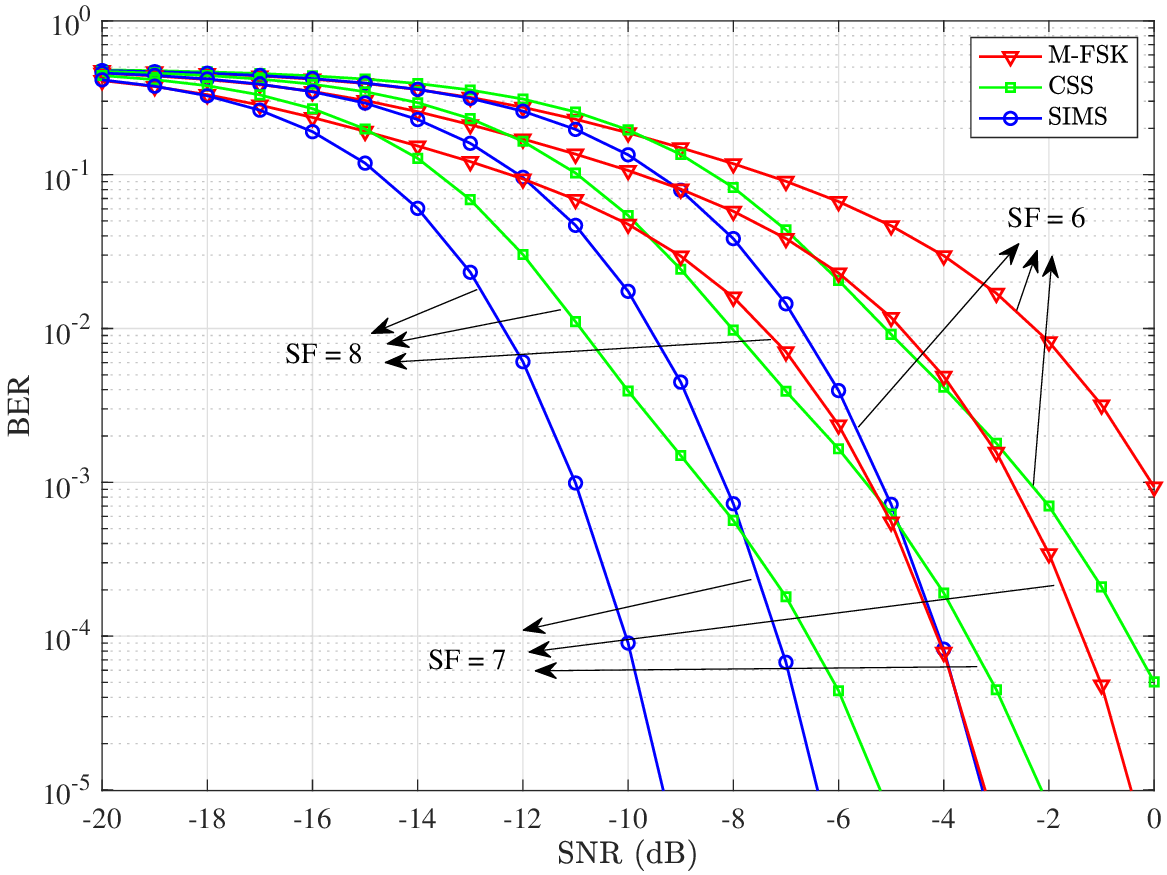}
	\caption{BER performance comparison of SIMS, CSS, and M-FSK schemes over frequency-selective channels without small-scale fading for different spreading factors.} 
	\label{Fig_BER_AWGN_2Paths}
	\vspace{-10pt}
\end{figure}

Fig.~\ref{Fig_BER_AWGN_2Paths} compares the BER of SIMS, CSS, and M-FSK in frequency-selective channels without small-scale fading. At $\text{BER}=10^{-4}$, CSS achieves about a $2$ \si{dB} gain over M-FSK, while SIMS provides an additional $3.5$ \si{dB} gain over CSS with the same SF. This result aligns with prior simulation studies, which show that CSS significantly outperforms M-FSK in frequency-selective channels due to the symbol-level frequency diversity introduced by chirps \cite{MARQUET2020595}. This advantage arises from the use of quasi-orthogonal sequences in SIMS.

This observation can be further explained by examining the correlation properties of the sequences. 
For an orthogonal sequence set $\{\bm{s}_i\}$, orthogonality holds strictly in frequency flat-fading channels: $\langle \bm{s}_i, \bm{s}_j \rangle = 0, \, i \neq j.$
However, in a frequency-selective channel with impulse response $h(n)$, the received signal is a convolution, and the effective correlation becomes $\langle \bm{s}_i * h, \, \bm{s}_j * h \rangle,$
which is generally nonzero due to delayed multipath components. Thus, orthogonality is destroyed and error performance degrades. 

In contrast, quasi-orthogonal sequences $\{\bm{q}_i\}$ are designed to maintain low cross-correlation under time shifts, i.e., $\big|\langle \bm{q}_i, \, \bm{q}_j(\tau)\rangle \big| \ll 1, \, \forall \tau$,
where $\bm{q}_j(\tau)$ denotes a delayed version of $\bm{q}_j$. This property ensures that even after multipath convolution, the correlation detector can still reliably distinguish different codewords, leading to the superior BER performance of SIMS.

\begin{figure}[!t]
	\centering
	\includegraphics[width=0.35\textwidth]{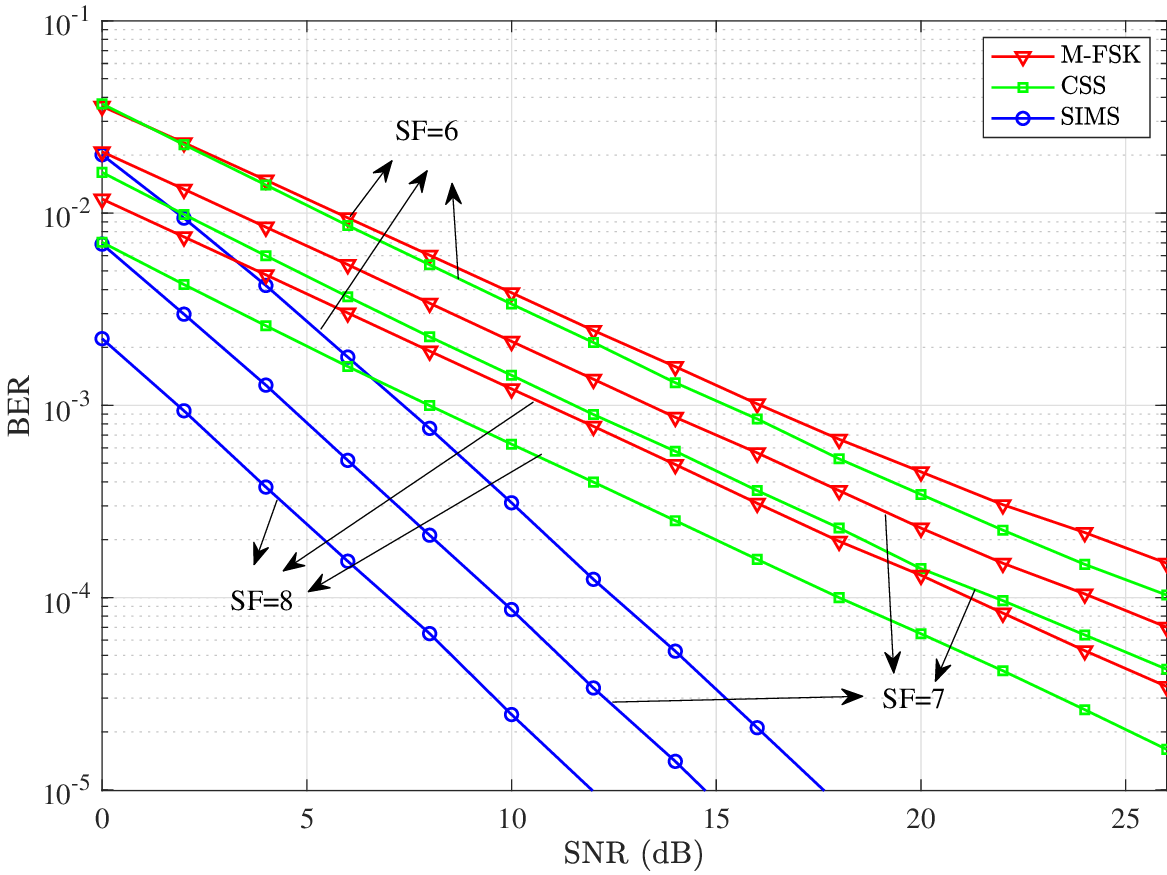}
	\caption{BER performance comparison of SIMS, CSS, and M-FSK schemes over frequency-selective channels with small-scale fading for different spreading factors.}
	\label{Fig_BER_Rayleigh_2Paths}
	\vspace{-10pt}
\end{figure}

Fig.~\ref{Fig_BER_Rayleigh_2Paths} compares BER performance in frequency-selective fading.  For the same SFs, CSS consistently outperforms M-FSK, with a larger gap at higher SFs. The performance difference between CSS and SIMS remains stable, with SIMS achieving about a $12$~\si{dB} gain at $\text{BER}=10^{-4}$ and $\text{SF}=7$. This gain highlights the strong multipath robustness of SIMS, which is attributed to its quasi-orthogonal code design, enabling reliable detection even under severe fading conditions.

In summary, SIMS demonstrates substantially better BER performance than both CSS and M-FSK in multipath fading environments, highlighting its strong potential as an enabling technology for future LPWANs.

\subsection{Multi-User Scenarios}
This subsection evaluates the robustness of MU-SIMS under asynchronous multi-user access. In the simulations, each user is assigned a distinct root sequence with favorable cross-correlation properties, and the aggregate system performance is measured by the average per-user BER. The results demonstrate the strong interference suppression capability of the SIMS codebooks and validate their scalability to large user populations.

\subsubsection{\ul{AWGN Channels}}

\begin{figure}[!t]
	\centering
	\includegraphics[width=0.35\textwidth]{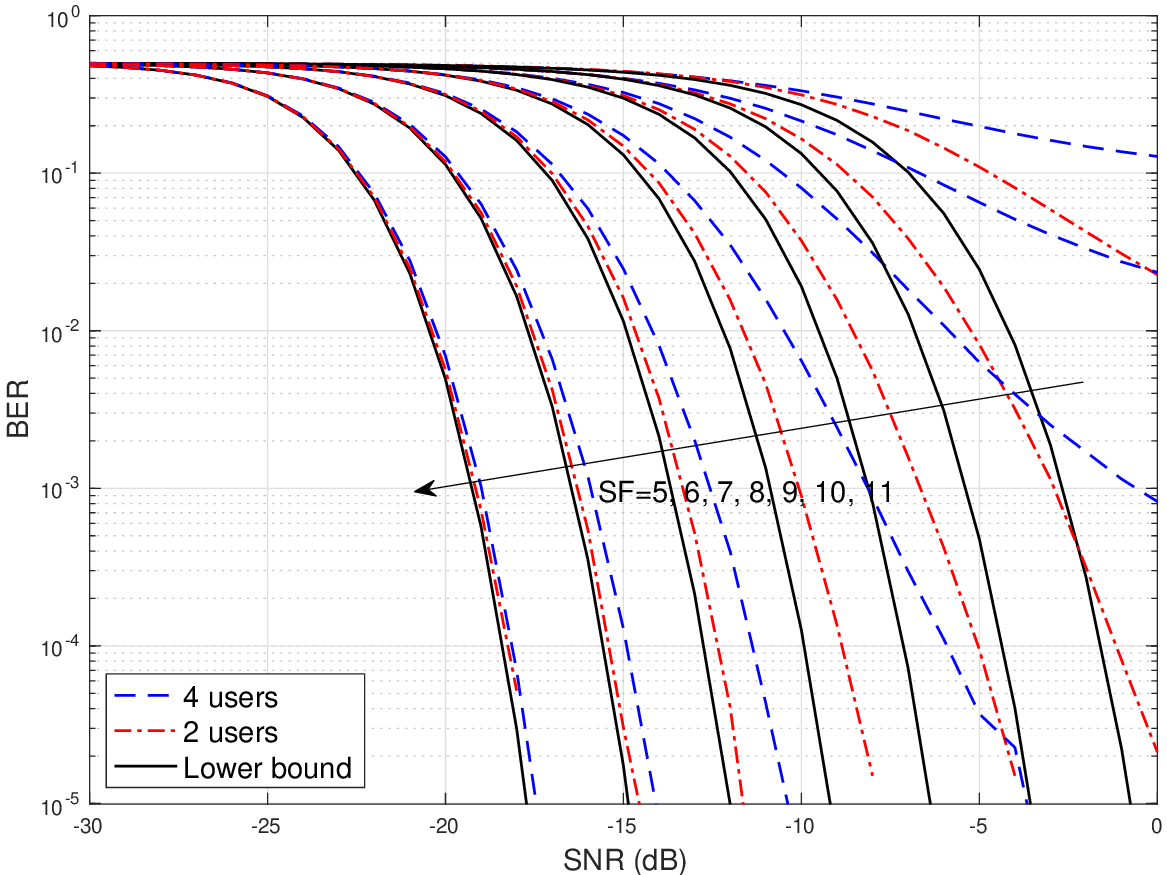}
	\caption{BER performance comparison of MU-SIMS over AWGN channels for varying spreading factors.}
	\label{Fig_BER_MUSIMS_AWGN}
	\vspace{-10pt}
\end{figure}

Fig.~\ref{Fig_BER_MUSIMS_AWGN} compares the BER performance of MU-SIMS over AWGN channels for different spreading factors (SFs). Unless otherwise specified, the reported BER represents the \emph{per-user average}. As SF increases, the simulated BER gradually approaches the theoretical single-user lower bound, indicating that inter-user interference is effectively suppressed by the quasi-orthogonal SIMS codebooks.

For small SF values (e.g., $\text{SF}=5$–$8$), the quasi-orthogonality between user codebooks is relatively weak, and the cross-correlation terms among user sequences become non-negligible, especially under asynchronous symbol timing. These residual correlations introduce mutual interference at the matched filter outputs, thereby increasing the detection error rate. Conversely, for large SF values (e.g., $\text{SF}=9$–$11$), the SIMS codebooks approach asymptotic orthogonality, as predicted by the probabilistic quasi-orthogonality theorem, and inter-user interference becomes negligible. Consequently, the BER curves converge toward the theoretical single-user limit.

Furthermore, the impact of user load is examined by varying the number of simultaneously active users, with two and four users considered in Fig.~\ref{Fig_BER_MUSIMS_AWGN}. As the number of users increases, the aggregate interference power grows approximately linearly, leading to a moderate degradation in BER. This degradation is more pronounced for small SF, where the limited spreading gain cannot fully decorrelate overlapping user symbols. In contrast, when SF is large, the quasi-orthogonality of the SIMS sequences ensures that the interference from other users behaves almost like additive white noise, and the per-user BER remains nearly invariant with the number of users.

From a theoretical perspective, this phenomenon aligns with the scaling law of quasi-orthogonal multi-user systems. As the sequence length $N=2^{\text{SF}}$ increases, the pairwise correlation magnitude among user codebooks decreases at a rate proportional to $\mathcal{O}(1/\sqrt{N})$ \cite[Eq.~(3.14)]{Vershynin2025}. Consequently, the residual multi-user interference diminishes with SF, verifying the asymptotic scalability of the SIMS framework.

\subsubsection{\ul{Rayleigh Fading Channels}}
In Rayleigh fading channels, similar trends are observed as in the AWGN case. The random amplitude and phase fluctuations introduced by fading partially decorrelate user sequences, thereby mitigating coherent multi-user interference at moderate SFs. However, the dominant factor governing performance remains the intrinsic quasi-orthogonality of the SIMS codebooks. As illustrated in Fig.~\ref{Fig_BER_MUSIMS_Rayleigh}, the BER degradation from the single-user to the multi-user case is negligible and further diminishes with increasing SF. For instance, when $\text{SF}=10$ or $11$, the BER curves for single-user and multi-user configurations are nearly identical.

\subsubsection{\ul{Frequency-Selective Fading Channels}}
\begin{figure}[!t]
	\centering
	\includegraphics[width=0.35\textwidth]{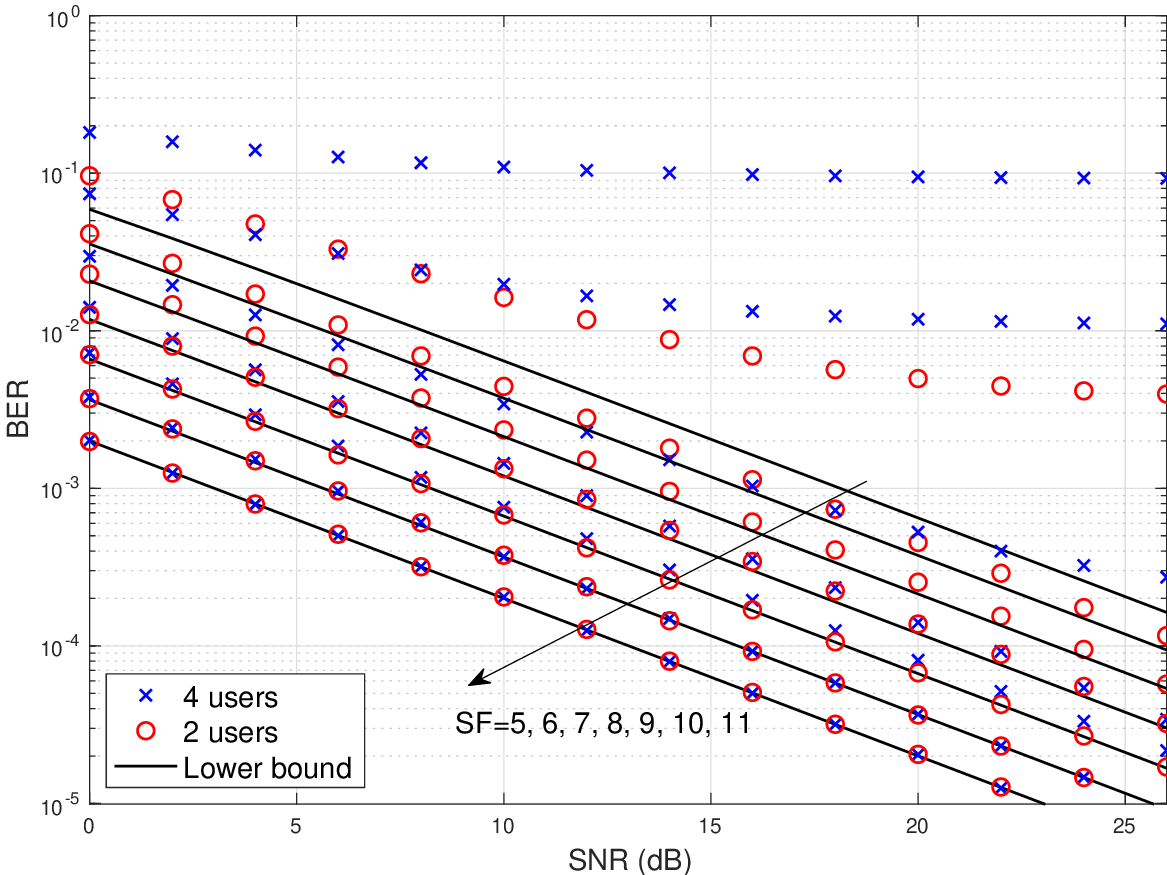}
	\caption{BER performance comparison of MU-SIMS over Rayleigh fading channels for varying spreading factors.}
	\label{Fig_BER_MUSIMS_Rayleigh}
	\vspace{-10pt}
\end{figure}

Fig.~\ref{Fig_BER_MUSIMS_Rayleigh_2Paths} presents the BER performance of the proposed multi-user SIMS scheme in frequency-selective Rayleigh fading channels under various SFs and user numbers. The observed trends are consistent with those in frequency-flat Rayleigh fading (Fig.~\ref{Fig_BER_MUSIMS_Rayleigh}), but with overall superior BER performance. This improvement stems from the frequency diversity gain inherent in multipath propagation, which can be effectively exploited using diversity combining techniques, such as maximum ratio combining, adopted in this study.

Moreover, Figs.~\ref{Fig_BER_MUSIMS_Rayleigh} and \ref{Fig_BER_MUSIMS_Rayleigh_2Paths} show that MU-SIMS performs robustly under asynchronous multi-user access, typical of LPWANs. Even with several-symbol misalignment, SIMS sequences maintain low cross-correlation, ensuring reliable detection. In contrast, conventional orthogonal codes, such as Walsh–Hadamard or Zadoff–Chu, degrade significantly under similar offsets, highlighting SIMS’s robustness to asynchrony.

\subsection{Complexity vs. Performance Tradeoff}

Table~\ref{Tab-Complexity-Comparison} summarizes the computational complexity of FSK, CSS, and SIMS across primary processing stages. FSK has the lowest complexity, with subcarrier selection and energy detection of order $\mathcal{O}(1)$ and $\mathcal{O}(K)$, respectively. CSS incurs a higher cost due to FFT/IFFT operations of $\mathcal{O}(N_{\text{SF}}\log N_{\text{SF}})$ for modulation and demodulation. The proposed SIMS scheme maintains comparable asymptotic complexity while introducing block-wise correlation and codebook mapping, resulting in $\mathcal{O}(KN_{\text{SF}}\log N_{\text{SF}})$ for multi-user detection. This moderate increase is offset by scalable codebooks and inherent parallelism, enabling efficient hardware implementation. Overall, SIMS achieves superior reliability and multi-user scalability with only marginal additional complexity over CSS.

\begin{figure}[!t]
	\centering
	\includegraphics[width=0.35\textwidth]{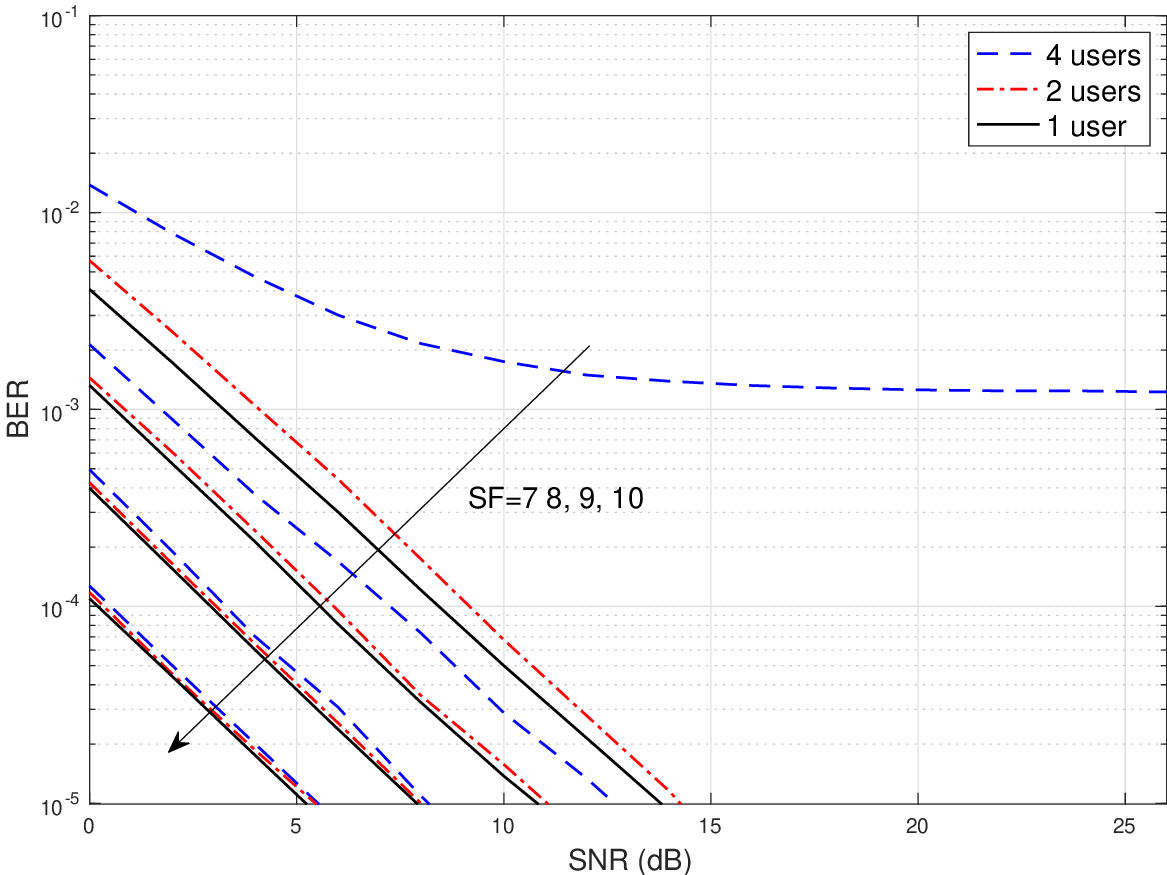}
	\caption{BER performance comparison of MU-SIMS over frequency-selective channels with small-scale Rayleigh fading for varying spreading factors.}
	\label{Fig_BER_MUSIMS_Rayleigh_2Paths}
	\vspace{-10pt}
\end{figure}

\begin{table*}[!t]
	\centering
	\caption{Computational Complexity Comparison of FSK, CSS, and SIMS Modulation Schemes}
	\label{Tab-Complexity-Comparison}
	\begin{tabular}{lccc}
		\toprule
		\textbf{Metric} & \textbf{FSK} & \textbf{CSS} & \textbf{SIMS} \\
		\midrule
		Modulation & $\mathcal{O}(1)$ per symbol & $\mathcal{O}(N_{\text{SF}}\log N_{\text{SF}})$ (IFFT) & $\mathcal{O}(N_{\text{SF}}\log N_{\text{SF}})$ (block transform) \\
		Demodulation & $\mathcal{O}(K)$ (energy detection) & $\mathcal{O}(N_{\text{SF}}\log N_{\text{SF}})$ (FFT) & $\mathcal{O}(K N_{\text{SF}}\log N_{\text{SF}})$ (multi-sequence correlation) \\
		Codebook & Fixed & Cyclic shifts & $\mathcal{O}(K N_{\text{SF}})$ (root-based) \\
		Memory & Low ($\mathcal{O}(K)$) & Moderate ($\mathcal{O}(N_{\text{SF}})$) & Medium--High ($\mathcal{O}(K N_{\text{SF}})$) \\
		Multi-user detection & Independent & Parallel FFTs & Joint correlation across users \\
		Scalability & Linear & Quasi-linear & Quasi-linear \\
		Parallelization & High & High & High \\
		Complexity–performance & Low / limited & Moderate / good & Higher / excellent \\
		\bottomrule
	\end{tabular}
\end{table*}

The results in Figs.~\ref{Fig_BER_MUSIMS_Rayleigh}–\ref{Fig_BER_MUSIMS_Rayleigh_2Paths} and Table~\ref{Tab-Complexity-Comparison} reveal a clear tradeoff between complexity and performance across the three modulation schemes. FSK offers the lowest complexity but limited spreading gain and poor multi-user scalability, making it suitable only for ultra-low-rate links. CSS enhances BER through cyclic spreading and FFT-based correlation, at moderate computational cost. The proposed SIMS scheme further exploits structured sequence diversity and block-wise processing, delivering superior reliability and robustness to asynchronous interference. Although its complexity scales as $\mathcal{O}(KN_{\text{SF}}\log N_{\text{SF}})$, the algorithm is highly parallelizable and efficiently implemented on GPU or FPGA platforms. Overall, SIMS achieves a favorable balance between complexity and performance, supporting near-orthogonal multi-user access with scalable detection capability.

\section{Concluding Remarks}
\label{Section-Conclusion}

This paper introduces a unified framework for block signal processing in LPWANs that overcomes the inherent limitations of conventional symbol-by-symbol transmission. By jointly incorporating the signal block vector, intra-block structure, and signal basis matrix, the framework enables systematic waveform design and low-complexity block-correlation demodulation. The probabilistic quasi-orthogonality of SIMS codewords, both within a single codebook and across multiple user codebooks, is also established, ensuring robust multi-user separability. In addition to its theoretical contributions, the framework naturally extends to practical waveforms such as FSK and CSS, offering a solid foundation for enhancing reliability, scalability, and physical-layer performance in future LPWAN systems.

\begin{appendix}[Proof of Lemma~\ref{Lemma-1}]
\label{Appendix-A}
For each chip index $n$, collect the $K$ entries across $k$:
\begin{equation}
    \bm{s}_l^{(n)} 
	=
    \begin{bmatrix}
        A(0) \exp\left(\mathrm{j} \tfrac{2\pi}{K} f_0 \vartheta c_l(n) 0 \right) \\
        A(1) \exp\left(\mathrm{j} \tfrac{2\pi}{K} f_0 \vartheta c_l(n) 1 \right) \\
        \vdots \\
        A(K{-}1) \exp\left(\mathrm{j} \tfrac{2\pi}{K} f_0 \vartheta c_l(n) (K-1) \right)
    \end{bmatrix}.
\end{equation}
This can be written compactly as
\begin{equation}
    \bm{s}_l^{(n)} = \bm{d}_l \odot \bm{g}_n
    = \operatorname{diag}(\bm{d}_l)\,\bm{g}_n,
\end{equation}
where $\bm{d}_l = \begin{bmatrix} A(0) & A(1) & \cdots & A(K-1) \end{bmatrix}^T$ denotes the symbol vector and $\bm{g}_n$ defined by \eqref{Eq-gn} represents the chip-dependent phase vector.

Stacking all chip blocks yields
\begin{equation}
    \bm{s}_l \;=\;
    \begin{bmatrix}
        \bm{s}_l^{(0)} \\ \bm{s}_l^{(1)} \\ \vdots \\ \bm{s}_l^{(N_\text{SF}-1)}
    \end{bmatrix}
    \;=\;
    \begin{bmatrix}
        \operatorname{diag}(\bm{d}_l)\,\bm{g}_0 \\
        \operatorname{diag}(\bm{d}_l)\,\bm{g}_1 \\
        \vdots \\
        \operatorname{diag}(\bm{d}_l)\,\bm{g}_{N_\text{SF}-1}
    \end{bmatrix}.
\end{equation}

Define the stacked spreading vector
\begin{equation}
    \bm{c}_l = \begin{bmatrix} \bm{g}_0^T & \bm{g}_1^T & \cdots & \bm{g}_{N_\text{SF}-1}^T \end{bmatrix}^T,
\end{equation}
so $\bm{c}_l \in \mathbb{C}^{K N_\text{SF}}$. The block-diagonal matrix $\bm{I}_{N_\text{SF}} \otimes \operatorname{diag}(\bm{d}_l)$
is a \((K N_\text{SF})\times(K N_\text{SF})\) matrix whose $n^{\mathrm{th}}$ diagonal block equals \(\operatorname{diag}(\bm{d}_l)\). Multiplying this block-diagonal matrix by $\bm{c}_l$ yields exactly the stacked vector above, which can be explicitly expressed as \eqref{eq:lemma1-main}. This completes the proof.
\end{appendix}

\bibliographystyle{IEEEtran}
\bibliography{References.bib}
\vfill

\end{document}